\documentclass{aa}
\usepackage{graphicx}
\usepackage{hyperref}
\usepackage{txfonts}
\usepackage{xcolor}
\usepackage{caption}
\usepackage{subcaption}
\usepackage{textcomp}
\usepackage[rightcaption]{sidecap}
\usepackage[mathlines]{linenoaa}

\graphicspath{{Figures/}}

\begin{document} 
\makeatletter
\let\linenumbers\relax
\let\modulolinenumbers\relax
\makeatother
	\title{
	Estimating early coronal mass ejection propagation direction with DIRECD during the severe May 8 and follow-up June 8, 2024 events}
\titlerunning{DIRECD for May 8 and June 8, 2024 CMEs}	
	
	\author{Shantanu Jain\inst{1}
		\and Tatiana Podladchikova \inst{1}
		\and Astrid M. Veronig\inst{2,3}
		\and Galina Chikunova\inst{4,1} 
		\and Karin Dissauer\inst{5}
		\and Mateja Dumbovic\inst{4}
		\and Amaia Razquin\inst{2}}
	
\institute{Skolkovo Institute of Science and Technology, Bolshoy Boulevard 30, bld. 1, 121205, Moscow, Russia \\
	\email{shantanu.jain@skoltech.ru}
	\and
	University of Graz, Institute of Physics, Universit\"atsplatz 5, 8010 Graz, Austria
	\and
	University of Graz, Kanzelh\"ohe Observatory for Solar and Environmental Research, Kanzelh\"ohe 19, 9521 Treffen, Austria
	\and
	University of Zagreb,  Faculty of Geodesy, Hvar Observatory, Kaciceva 26, HR-10000, Zagreb, Croatia
	\and
	NorthWest Research Associates, 3380 Mitchell Lane, Boulder 80301, CO, USA}
	
	\date{Received Month, Year; accepted Month, year}

	\abstract
	{On May 8, 2024, solar active region 13664 produced an X-class flare, several M-class flares, and multiple Coronal Mass Ejections (CMEs) directed towards Earth. The initial CME resulted in coronal dimmings, characterized by localized reductions in extreme-ultraviolet (EUV) emissions, indicative of mass loss and expansion during the eruption. On June 8, 2024 after one solar rotation, the same active region produced another eruptive M-class flare followed by coronal dimmings observed by the SDO and STEREO spacecraft.}
	{We analyse the early CME evolution and propagation direction from the expansion of the coronal dimming observed low in the corona using the DIRECD (Dimming Inferred Estimation of CME Direction) method.}
	{The DIRECD method derives key parameters of early CME propagation from the expansion behavior of the associated coronal dimming at the end of its impulsive phase by generating a 3D CME cone model whose orthogonal projection on the solar sphere matches the dimming geometry. To validate the resulting 3D CME cone, we compare the CME properties derived in the low corona with white-light coronagraph data.}
	{Using DIRECD, we find that the CME on 8 May 2024 expands close to radially, with an inclination angle of 7.7$^\circ$, an angular width of $70^\circ$, and a cone height of $0.81~R_{\text{sun}}$, which was derived at the end of the dimming's impulsive phase, and for which the CME shows connections to the dimming and still leaves footprints in the low corona. It was inclined 7.6$^\circ$ to the North in the meridional plane and 1.1$^\circ$ to the East in the equatorial plane. The CME on 8 June 2024, after one solar rotation, is inclined by 15.7$^\circ$ from the radial direction, has an angular width of $81^\circ$, and a cone height of $0.89~R_{\text{sun}}$. 
    The CME was inclined 6.9$^\circ$ to the South in the meridional plane and 14.9$^\circ$ to the West in the equatorial plane. Validation with white-light coronagraph data confirmed the 3D cone's accuracy by matching CME characteristics and projections with STEREO-A COR2 observations.}
	{Our study demonstrates that tracking low coronal signatures such as coronal dimming expansion in 2D for the May and June 2024 CMEs can estimate the 3D CME direction early in its evolution providing early lead times for mitigating adverse space weather impacts.}

	
	\authorrunning{S. Jain et al.}
	\maketitle
	

\section{Introduction}
Coronal Mass Ejections (CMEs) are the most powerful events in the solar system, causing significant disturbances to space weather. These phenomena involve clouds of magnetized plasma being expelled from the Sun at very high velocities \citep{michalek2009expansion, gopalswamy2009soho, tsurutani2014extreme, Cheng2017, Veronig2018_Genesis}. CMEs can pose a threat to technological systems, as they can disrupt radio transmissions, induce currents in power grids, and lower the orbits of Earth-orbiting satellites \citep{sandford1999impact,doherty2004space,baker2013major}. 
Statistical analysis shows that fast CMEs launched sequentially from the same active region are more geoeffective than isolated CMEs \citep{vennerstrom2016extreme,lugaz2017interaction, gomez2020clustering,scolini2020cme}. Thus it is important to detect and study early evolution of CMEs. Since early CME evolution is not well observed from coronagraphs due to projection effects \citep{schwenn2005association} and early location cannot be known due to occulation disk,  
a number of studies have tried to explore whether any CME-related phenomena may give us more insight into early CME evolution.
	
Coronal dimmings are one such phenomena associated with CMEs. They are regions of reduced extreme ultraviolet (EUV) and X-ray emissions in the low corona \citep{hudson1996long,sterling1997yohkoh,thompson1998soho}. Numerous studies have explored the strong connection between the initial development of CMEs and coronal dimmings in the lower corona. Some of these investigations examined how coronal dimmings relate to the mass of CME(e.g. \citep{harrison2000spectroscopic, harrison2003coronal, zhukov2004nature, lopez2017mass}, 
the morphology and early evolution of the CME \citep{attrill2006using, qiu2017gradual}, and their timing \citep{miklenic2011coronal, ronca2024recoverycoronaldimmings}. Additionally, various studies have examined the statistical relationship between CMEs and coronal dimmings \citep{bewsher2008relationship, reinard2009relationship, mason2016relationship, krista2017statistical, aschwanden2017global}, revealing strong correlations between CME mass and dimming area,  dimming brightness and the magnetic flux within the dimming region , as well as the CME's maximum speed and parameters such as area growth rate, brightness change rate, and magnetic flux rate, along with the mean intensity of the dimmings \citep{Dissauer2018b, Dissauer2019,chikunova2020coronal}. 

Several models exist to determine CME propagation direction and deflections. One such model is the Graduated Cylindrical Shell (GCS) model \citep{thernisien2006modeling, thernisien2011implementation}, which is a parametric, forward-fitting approach used to describe the three-dimensional shape of CMEs based on stereoscopic observations from coronagraphs at heights of up to 20 Rsun. Another method is elliptical tie-pointing \citep{byrne2010propagation}, which involves tracing the CME’s elliptical front from multiple viewpoints (e.g., from two spacecraft like STEREO-A and STEREO-B) and determining its 3D structure by triangulating the positions of corresponding points in each view. Additionally, geometric triangulation \citep{liu2010reconstructing, Podladchikova2019} is used to estimate the 3D position of a CME by measuring the same feature in images taken from two different viewpoints and using the angular separation between the spacecraft to infer the 3D position of specific points or features within the CME structure. These and other techniques contribute to understanding CME propagation in space.

\citep{Chikunova2023} found a link between dimming morphology and the 3D structure of CMEs, suggesting that dimming observations might also provide insights into early CME evolution. This idea was further explored using the DIRECD method described by \citet{jain2024coronal} to infer the initial CME propagation direction from coronal dimmings. The DIRECD method focuses on early-stage CME direction estimation using coronal dimming, while the GCS and other models provide a more detailed 3D structure of CMEs after they are observed in coronagraphs.

In this study, we analyse two recent extreme space weather events that occurred on 8 May 2024 and 8 June 2024, when CMEs erupted from the same active region AR3664 (AR3697 after one solar rotation), and estimate CME parameters in its early evolution from the associated coronal dimmings using the DIRECD method,  with its selection criteria further refined.
The associated flares were listed on XRT flare catalog \citep{watanabe2012hinode} and the publicly available DONKI\footnote{\url{https://ccmc.gsfc.nasa.gov/tools/DONKI/}} (Space Weather Database Of Notifications, Knowledge, Information) catalog. During the May 2024 storms, agencies reported a variety of impacts such as satellites experienced increased drag \citep{parker2024satellite} and strong geomagnetic disturbances in the earth's atmosphere \citep{hayakawa2024solar}.


\section{Event of May 8, 2024}

In the early hours of May 8, 2024 a series of eruptions were ejected into interplanetary space from active region 13364 (one of the largest in the current solar cycle). We focus our study to the first CME that erupted at 04:30~UT followed by three CMEs close after (see related video using Jhelioviewer \citep{muller2017jhelioviewer}). A total of six CMEs erupted 
that arrived near Earth on May 10. At around 12:30~UT on May 10, the first CME reached Earth, leading to a significant geomagnetic storm including auroras as far as 21$^\circ$ latitude \citep{parker2024satellite} and increased ionospheric activity \citep{spogli2024effects}. Moreover, around 5000 spacecrafts had to perform altitude adjustments to counteract the effects of a geomagnetic storm \citep{parker2024satellite}.
The Dst index, which measures the intensity of geomagnetic storms and is primarily linked to the magnetospheric ring current \citep{Gonzalez1994}, peaked at -412~nT (real-time quicklook Dst index) on May 11, 2024, at 03:00~UT. This amplitude was accurately predicted several hours in advance by the StormFocus service\footnote{\url{https://spaceweather.ru/content/extended-geomagnetic-storm-forecast}}, as shown in the prediction screenshot\footnote{\url{http://www.iki.rssi.ru/forecast/data/Archive/2024/en/05/11/Prediction_20240511_0805.gif}}. The forecast was based on real-time solar wind and interplanetary magnetic field data provided at the L1 Sun-Earth libration point \citep{Podladchikova2012,Podladchikova2018}. This was the largest geomagnetic storm since Nov 20, 2003, with a peak Dst of -422~nT, and since 1995 records, no storms have surpassed this intensity. This makes the May 8, 2024 event the second strongest storm of the 21st century. 

The first solar eruptions in the early hours of 8 May 2024 \citep{hayakawa2024solar} were followed by coronal dimming observed on solar surface by Solar Dynamics Observatory \citep[SDO]{pesnell2012solar} and Solar Terrestrial Relations Observatory \citep[STEREO]{kaiser2008stereo} spacecraft. In this 
study, we will use the information from these dimming observed by SDO to estimate CME properties such as 
angular width, inclination from the radial direction, and height, where the CME connects to the dimming and relate it with white-light coronagraph data.

\subsection{Dimming detection for the May 8, 2024 event} \label{may_preparation}
	
We detect coronal dimmings using the region-growing algorithm described by \cite{Dissauer2018a,Dissauer2018b,Dissauer2019} from a series of 193~\AA~images (4096 $\times$ 4096 pixel) from the Atmospheric Imaging Assembly \citep[AIA]{lemen2012atmospheric} telescopes aboard the Solar Dynamics Observatory - SDO spacecraft with a 1-minute cadence within the time range 04:30 - 06:30~UT.  We take a pre-eruption base map (30 minutes prior to the start time, at 04:00~UT) and apply differential rotation correction to the same reference time (04:00~UT) for all the images in the sequence to compare the drop in EUV intensity with respect to the pre-eruptive coronal state. 
The SDO/AIA images were processed and calibrated using the aiapy programs in the SunPy library
\citep{sunpy_community2020}. We further cropped the images around the eruption source determined by visual analysis from JHelioviewer client, focusing on a region of 1000$\times$1000 arcseconds. The dimming detection algorithm was applied to logarithmic base-ratio images using the automatic thresholding technique \citep{Dissauer2018a}. We extracted the darkest 30\% of pixels with a logarithmic threshold of $-0.15$ and applied median filtering with a box of 10$\times$10 pixels to reduce noise in the images. The detected dimmings pixels were saved in a cumulative map at each time step, and the area of dimming expansion was calculated using the method presented in \citet{Chikunova2023}, which allows us to estimate the surface area of a sphere for each pixel\footnote{\url{https://github.com/Chigaga/area_calculation}}.

\begin{figure}[h]
	\centering
	\includegraphics[width=0.5\textwidth]{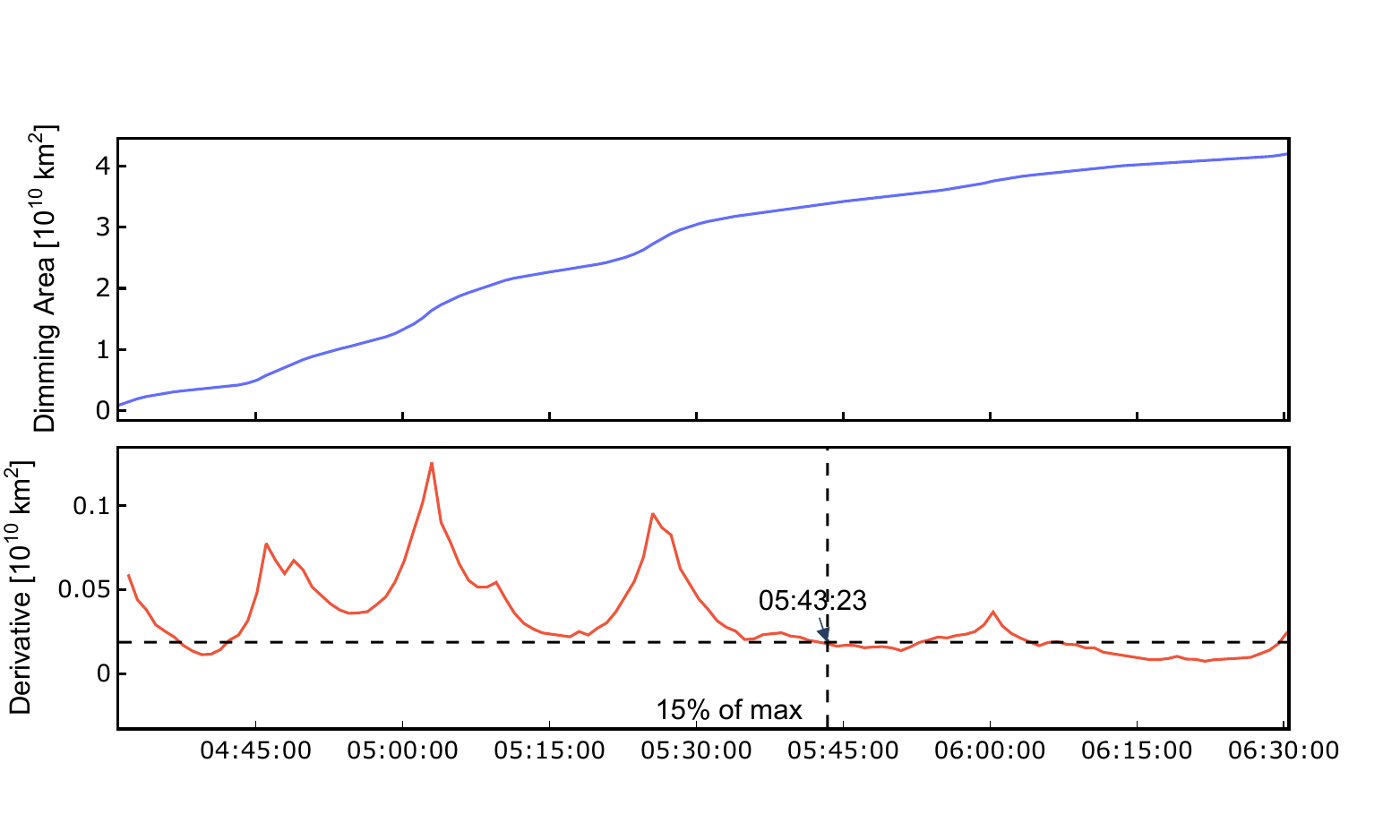}
	\caption{Expansion of dimming area $A_{t}$ (top panel) and its time derivative $dA_{t}/dt$ (bottom panel) over 2 hours for the 8 May, 2024 event. The end of the impulsive phase(indicated by the vertical dashed line) is defined as the time when the derivative of the dimming area curve, $dA_{t}/dt,$ has declined back to 15\% of its maximum value.}
	\label{area_der_may}
\end{figure}

Figure~\ref{area_der_may} presents the evolution of the derived dimming area, $A_{t}$ (top panel) and its time derivative $dA/dt$ (bottom panel) over a duration of 2 hours. We determine the end of the impulsive phase to be 05:43~UT, defined by the criteria in \citep{Dissauer2018a}, where the end of the impulsive phase occurs when the change rate of the dimming area, $dA/dt$, falls below 15\% of its maximum value.
	
\begin{figure}
	\centering
	\includegraphics[width=0.5\textwidth]{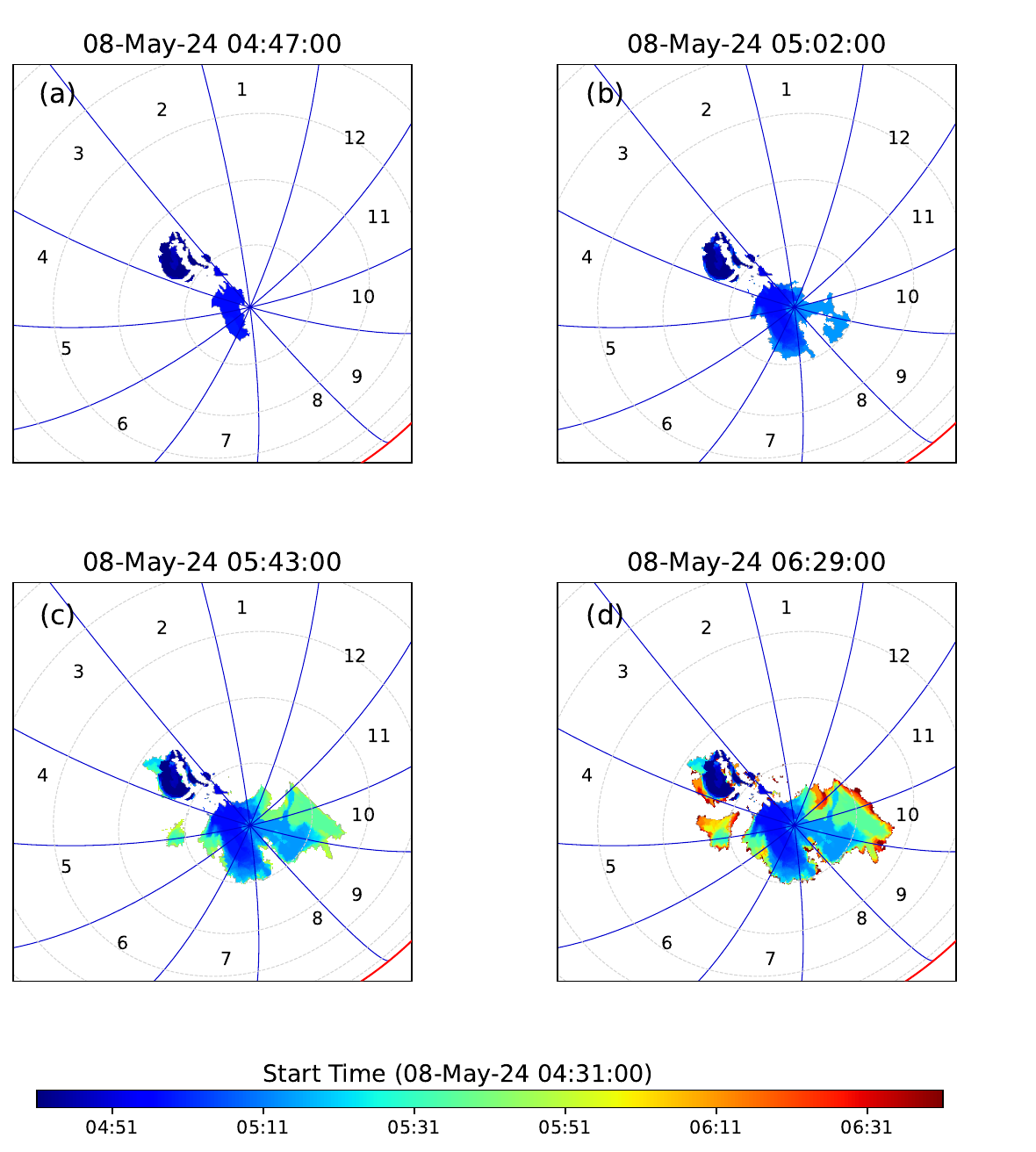}
	\caption{Dimming evolution for the 8 May 2024 CME at four different time steps: (a) 15 minutes before the maximum of the impulsive phase (04:47 UT), (b) at the max of the impulsive phase (05:02 UT), (c) at the end of the impulsive phase (05:43 UT) and (d) 2 hours after the start. The blue lines indicate 12 angular sectors and the colour bar shows when each dimming pixel was detected for the first time.}
	\label{dimming_expansion_may}
\end{figure}

Figure~\ref{dimming_expansion_may} shows the timing map of dimming evolution. Each dimming pixel is coloured according to the time it was first detected relative to the event start time. Panels (a,b) show cumulative dimming maps 15 minutes before and at the max of the impulsive phase 04:47 UT and 05:02 UT respectively. Panel (c) shows the dimming map at the end of impulsive phase (05:43~UT) and panel (d) shows the dimming map at the end of 
considered time period (06:29~UT). For DIRECD, we will use the dimming map 
at the end of the impulsive phase (panel c).

\subsection{Application of DIRECD to the May 2024 event}


The DIRECD method to estimate the early propagation of CMEs from the expansion of dimmings involves the following main steps. First, we estimate the dominant direction of the dimming extent based on the evolution of the dimming area (Fig. \ref{dimming_expansion_may}). Second, using the derived dominant direction of the dimming evolution on the solar sphere, we solve an inverse problem to reconstruct an ensemble of 3D CME cones at different heights, widths, and deflections from the radial propagation. Third, we search for the CME parameters for which the 3D cone orthogonal projections onto the solar sphere would match the geometry of the dimming at the end of its impulsive phase best.

\begin{figure}
	\begin{subfigure}{\columnwidth}
		\centering
		\includegraphics[width=0.7\textwidth]{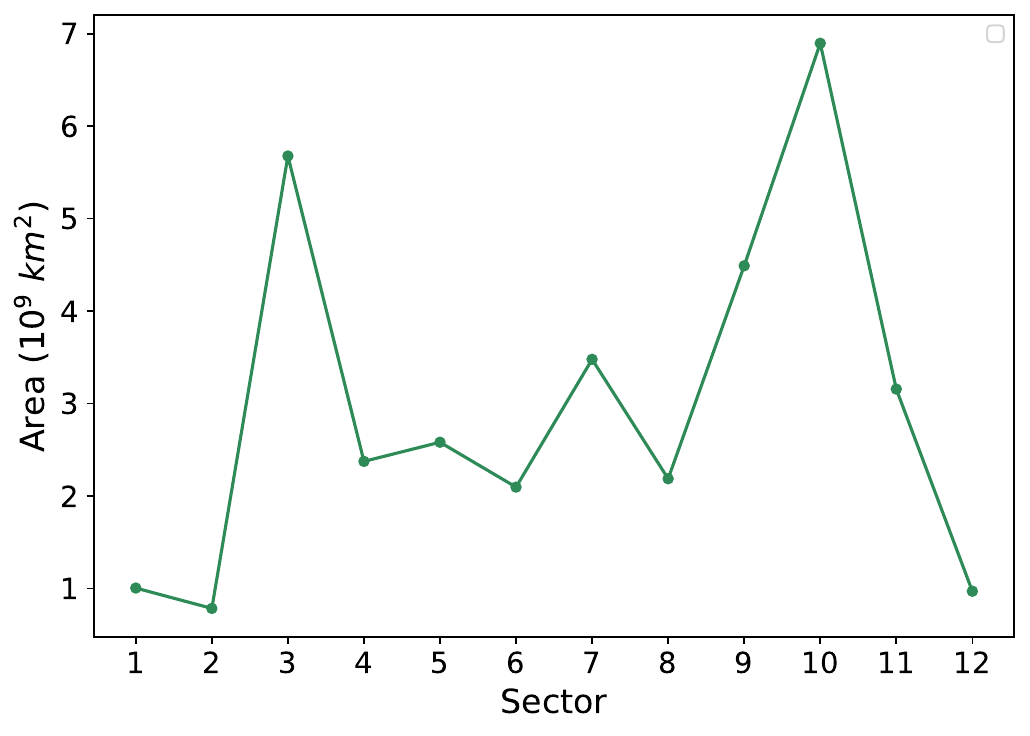}
		\caption{}
		\label{may_area_sector_end}
	\end{subfigure}    
	\begin{subfigure}{\columnwidth}    
	\includegraphics[width=0.9\textwidth]{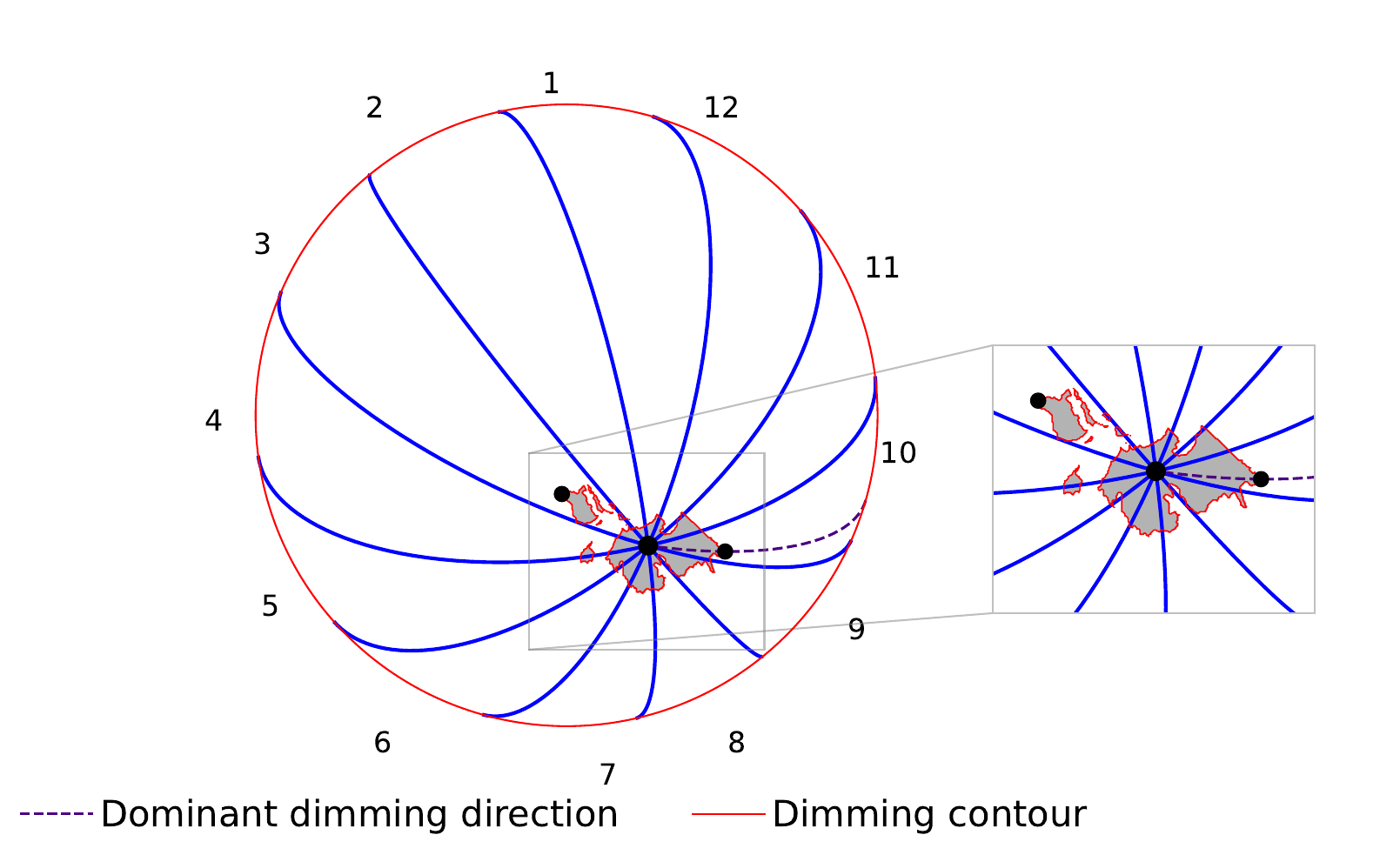}
		\caption{}
		\label{dominant_sector_may_end}  
	\end{subfigure}
	\caption{Dominant direction of dimming expansion at the end of the impulsive phase for 8 May 2024. Top: Dimming area in the different sectors (shown in the bottom panels in blue), revealing its maximum	extent in sector 10. Bottom: cumulative dimming pixel mask outlined by gray with red contours. The blue lines indicate the 12 sectors. The dashed purple line gives the sector of dominant dimming development. The small black dots show the dimming edges at the farthest edges from the source (one of them being in the sector of dominant dimming direction and the other being in sector 3) which are used to generate the 3D CME cones at different heights, associated widths and inclination angles. }

\end{figure}

Figure~\ref{may_area_sector_end} shows the area distribution curve for each sector at 
the end of the dimming impulsive phase.
We define the dominant dimming direction in the middle of the sector with the largest dimming area (sector 10).
Further, we choose two dimming edges which are farthest from the source which is the centroid of the detected dimming. Here one of them is in sector 10 along the dominant dimming direction and the other is in sector 3 as shown in Figure~\ref{dominant_sector_may_end}. We then construct two 3D lines from the Sun's centre to the two dimming edges. These lines form the boundaries of an ensemble of 3D cones, whose projections align with the dimming geometry.  We then generate an ensemble of 3D CME cones where every cone is defined by its unique height, inclination angle, and width, maintaining a connection to the dimming. Given the relationship between CMEs and dimmings, the CME cone naturally tilts towards the dominant dimming direction, aligning the axis of the smaller cone with this direction. As the inclination angle increases, the length of the main cone axis shortens. In our approach, we fix the small cone axis while allowing the larger cone axis to vary, making the identification of the dominant dimming direction a critical and advantageous part of the method.

Figure~\ref{may_projections} in Appendix~ \ref{Appendix_A} shows an ensemble of twenty such cones with heights ranging from 0.15-2.08~$R_{sun}$, angular widths varying from $132.6-49.8^\circ$ and changing inclination angles of $24.4-5.2^\circ$ (Cols. 1 and 3) and their orthogonal projections onto the solar sphere (Cols. 2 and 4). Columns 1 and 3 depict the top view to better show the reconstructed cones, and columns 2 and 4 show the face-on view. The heights in the constructed ensemble of cones is a representation of the heights at which the CME is assumed to leave footprints in the low corona and remains connected to the dimming. As can be seen from figure \ref{may_projections}, the cone projections are required to be wider for smaller heights and become narrower for larger heights albeit marginally to fit the observed dimming extent.

\begin{figure}
	\centering
	\includegraphics[width=0.5\textwidth]{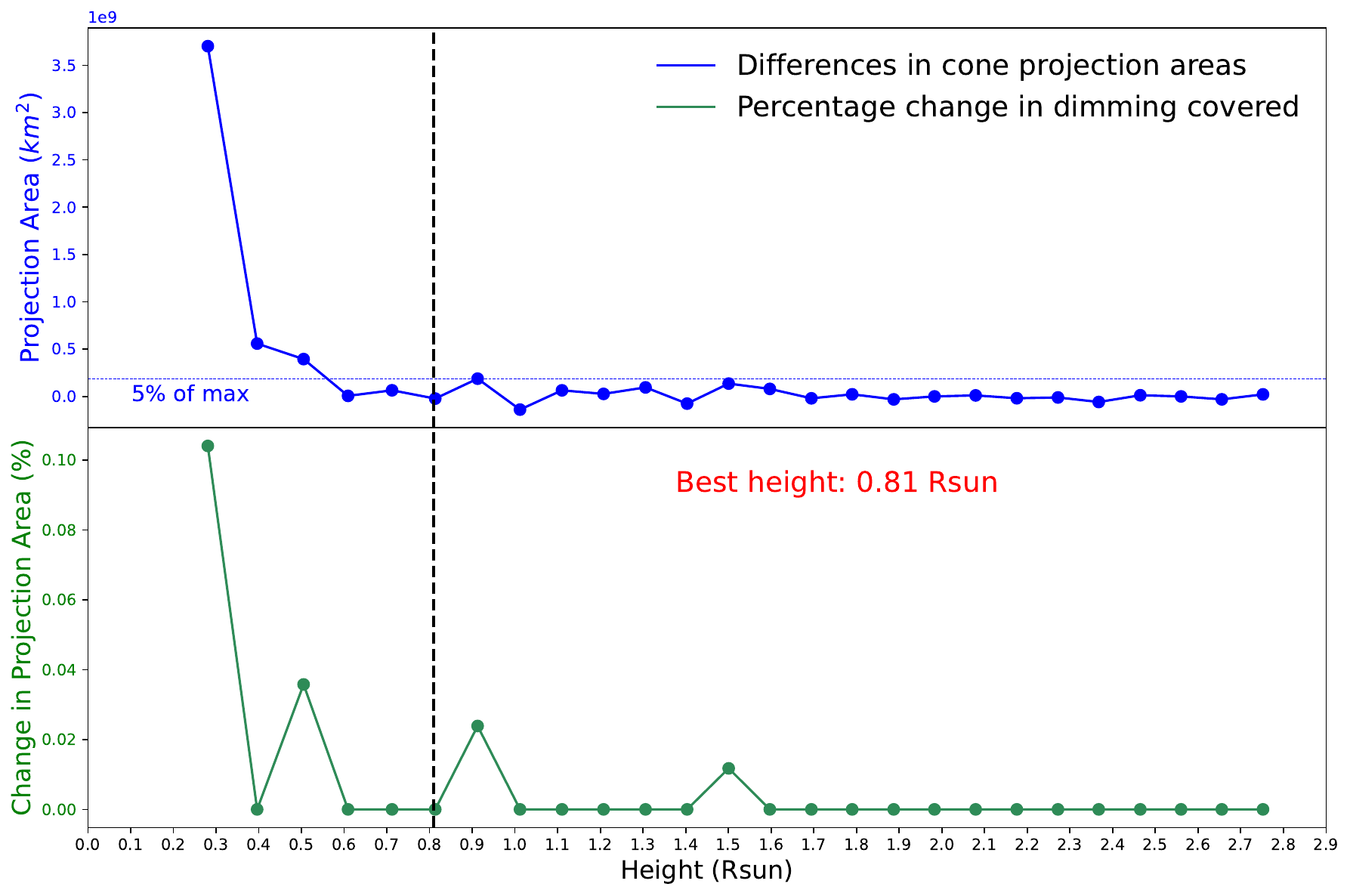}
	\caption{
		Consecutive differences of cone projection areas 	$A_{\Delta h_{1,2}}$ defined by Equation~\ref{criteria_1} (blue line, top panel) and the percentage of relative change of the dimming area inside the projection area for consecutive cones (green line, bottom panel) for the May 8, 2024 event. The vertical dashed line indicates the step (associated with a cone height of 0.81~$\rm{R_{sun}}$, width of 69.8$^\circ$ and inclination angle of 7.7$^\circ$) where consecutive differences (blue line) reach 5\% of the maximum of the cone projection area and just before the increase in percentage of relative change of the dimming area within projections (green line).
	}
	\label{best_height_may_1}
\end{figure}

\begin{figure}
	\captionsetup[subfigure]{labelformat=empty}	
	\begin{subfigure}{0.45\columnwidth}
		\centering
    \includegraphics[width=\textwidth]{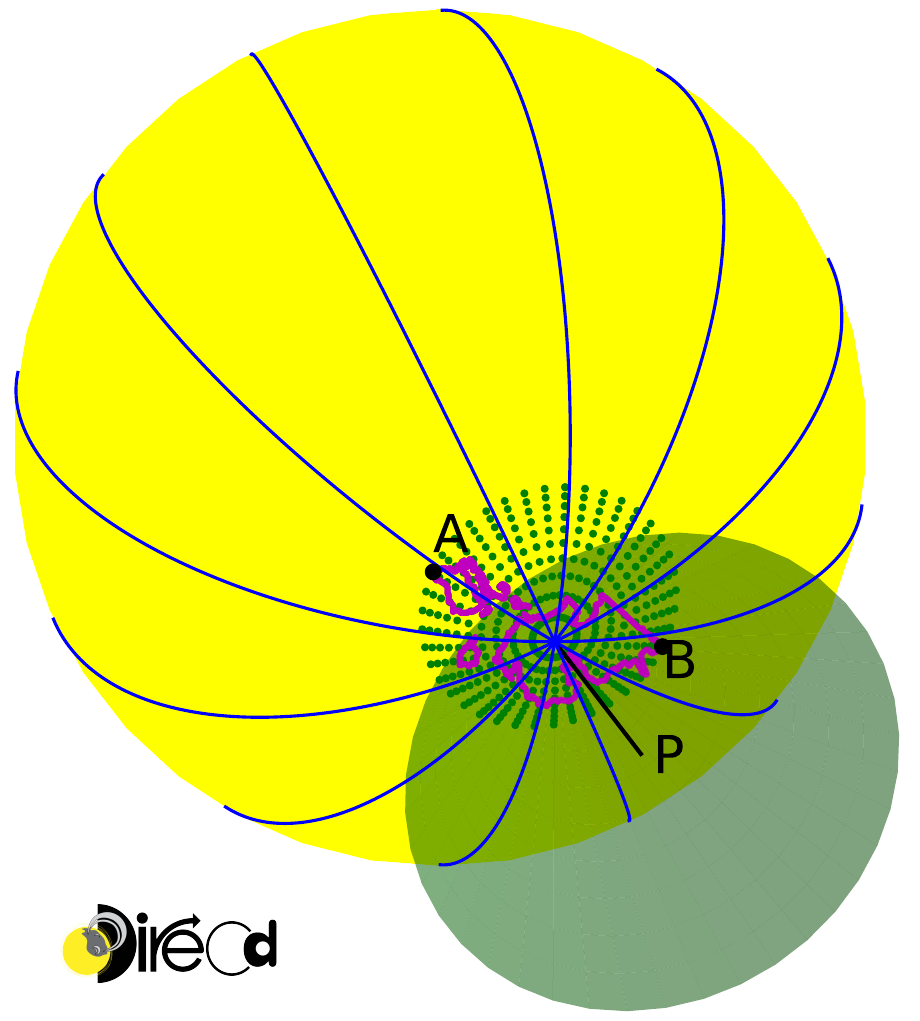}
		\caption{(a)}
		\label{best_fit_may}
	\end{subfigure}    
	\begin{subfigure}{0.5\columnwidth}    
		\centering
		\includegraphics[width=\textwidth]
		{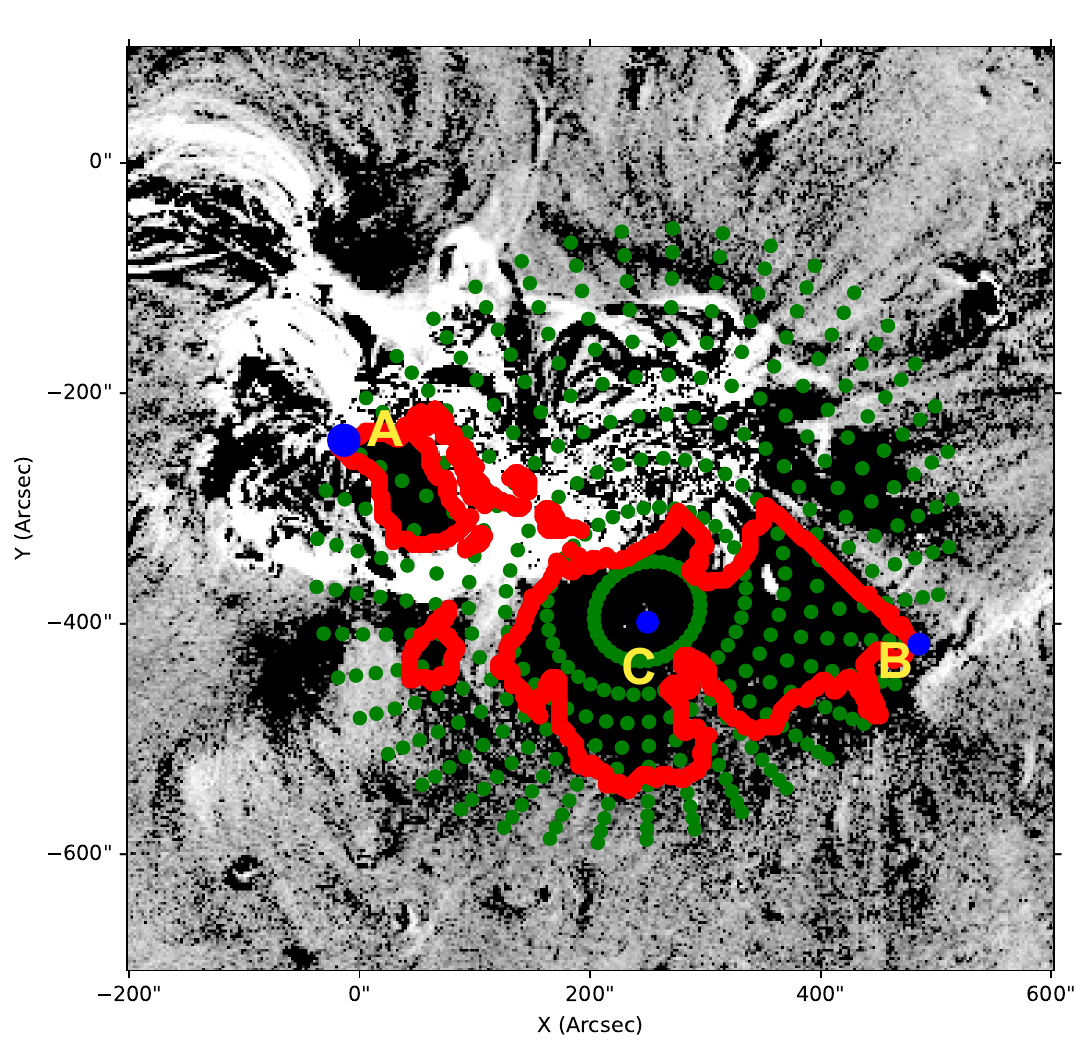}
		\caption{(b)}

		\label{SDO_may}  
	\end{subfigure}
	\caption{(a) Application of the DIRECD method to relate the expansion of coronal dimming to the early CME propagation. Magenta boundaries outline the dimming. Points A and B mark the largest North and South dimming extent. The best-fit green cone has a height of 0.81 Rsun, width of 69.8$^\circ$ and inclination angle of . Green dots indicate the orthogonal projections of the CME cone onto the solar surface. We require an edge of the cone base to be orthogonally projected to points A and B to match the dimming extent. (b) Dimming detection: 193 Å SDO/AIA base-difference image together with the boundary of the identified dimming region (red) at the end of impulsive phase	05:43 UT. Points A and B mark the largest North and South dimming extent.}
	\label{}
\end{figure}

%

To quantify the dynamics of the cone projections and to choose the 3D CME cone, which matches best the dimming geometry, we first consider the consecutive differences of cone projection areas 
$A_{\Delta h_{i,j}}$ for every cone height ($h_{i}, h_{j}$)
\begin{equation}\label{criteria_1}
	A_{\Delta h_{i,j}} = A_{h_{j}} - A_{h_{i}}
\end{equation}
The evolution of $A_{\Delta h_{1,2}}$ is shown in Figure~\ref{best_height_may_1} (blue line, top panel) and indicates the projection areas shrinking.  We select the cone that best fits the dimming at the point of the most shrinking in projection areas, where the cone projections remain sensitive to changes in cone parameters, estimated at 5\% (vertical dashed line) of the blue line peak. In the current case, this corresponds to the 3D CME cone with a height range of 0.61~$\rm{R_{sun}}$ -- 0.81~$\rm{R_{sun}}$, corresponding to a width range of 79.0$^\circ$ -- 69.8$^\circ$ and inclination angles of 9.0$^\circ$ -- 7.7$^\circ$. 

Additionally, we require the majority of the dimming area to be inside the cone projection by considering the percentage of relative change of the dimming area inside the projection area for the consecutive cones 
(Figure~\ref{best_height_may_1}, green line, bottom panel) defined as:
\begin{equation} \label{criteria_2}
\frac{A_{d}}{A_{p}} = \frac{S_{k} - S_{k+1}}{S_{k}},
\end{equation}
where $A_{d}$ and $A_{p}$ are the dimming and projection area respectively. $S_{k}$ is the area of dimming inside the projection area for cone height $h_{k}$, $S_{k+1}$ is the area of dimming inside the projection area for the consecutive cone height $h_{k+1}$. 

We choose the 3D CME cone that best matches the dimming just before the abrupt increase of the criteria above, which is in range of the initial choice of the CME parameters early in its evolution (see details in Appendix~\ref{Appendix_B}). Figure \ref{best_fit_may} shows the resulting best-fit 3D CME cone (green) with a height of 0.81~$\rm{R_{sun}}$, width of 69.8$^\circ$ and inclination angle of 7.7$^\circ$ and its orthogonal projections (green dots) together with the dimming boundaries at the end of its impulsive phase. 

The height of 0.81~$\rm{R_{sun}}$ here indicates the height up to which the CME is assumed to leave footprints in the low corona and remains connected to the dimming \citep{jain2024coronal}. Figure \ref{SDO_may} shows a 193Å SDO/AIA base-difference image, along with the boundary of the identified dimming region (in red) at the end of the impulsive phase. The source is marked by point C. The largest extent of
the dimming in the North and South direction is indicated by points A and B, respectively.

\begin{figure}  
	\centering      
	\includegraphics[width=0.3\textwidth] {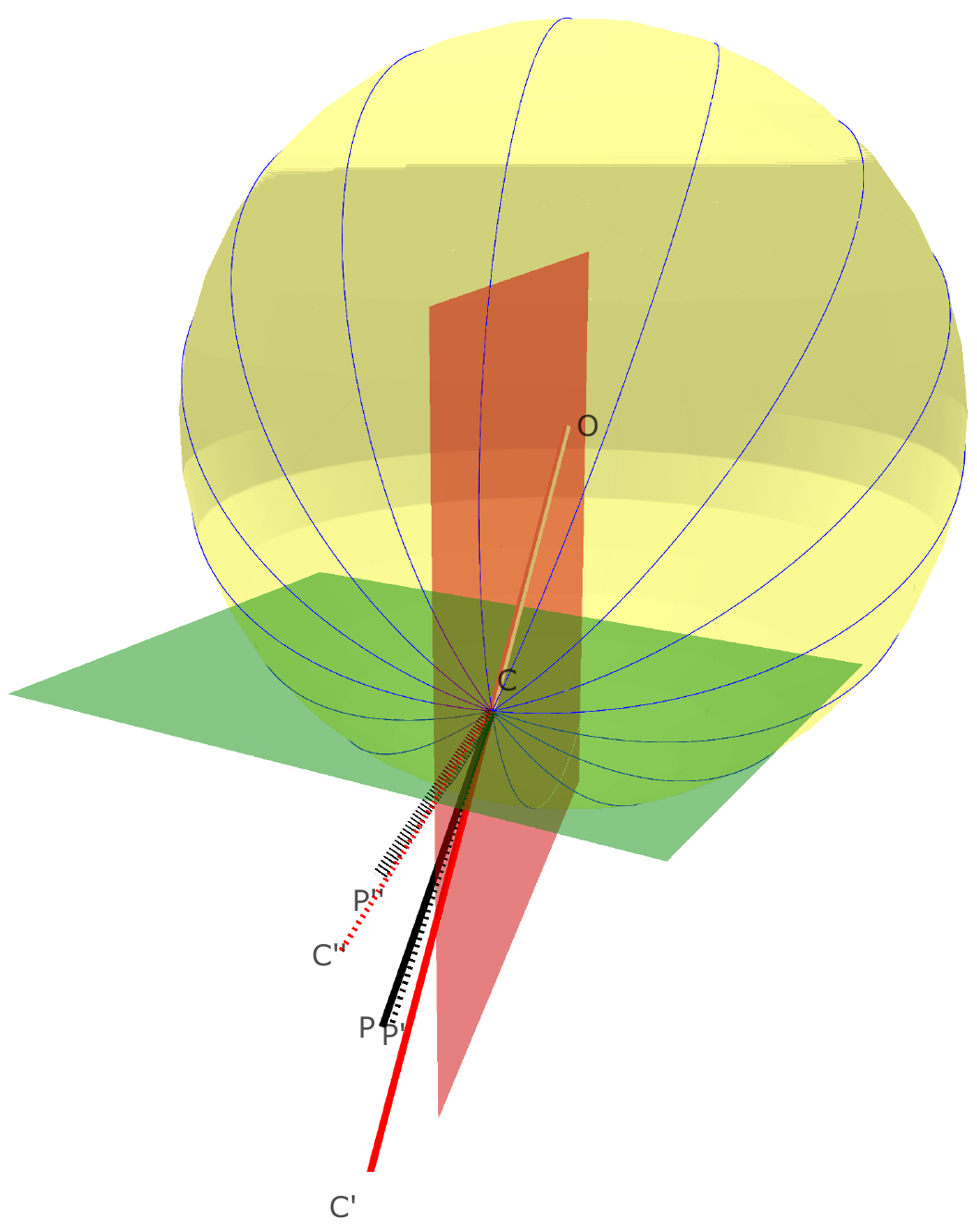}
	\caption{Projection of the CME direction onto the meridional (red) and equatorial (green) planes. The radial direction lies in the red meridional plane by definition (along the line CC'), the
		dashed red line shows its projection onto the equatorial plane (CC''). The dashed black line gives the projection of the cone central axis
		CP (the CME direction in 3D) to the meridional plane (CP'), and the equatorial plane (CP'').}

	\label{planes_may}
\end{figure}

Figure \ref{planes_may} shows the best-fit cone inclination within the radial and meridional planes.  The red meridional plane intersects the Sun's centre O, the source C on the solar surface, and the North Pole. The green plane passes through point C and is aligned parallel to the Sun's equatorial plane.

We obtain that the 8 May 2024 is inclined from the radial direction by 7.6 $^\circ$  to the North (i.e, in the
meridional plane) and 1.1$^\circ$  to the East (in the equatorial plane).

\begin{figure}
	\captionsetup[subfigure]{labelformat=empty}	
	\begin{subfigure}{0.49\columnwidth}
		\centering
		\includegraphics[width=\textwidth]{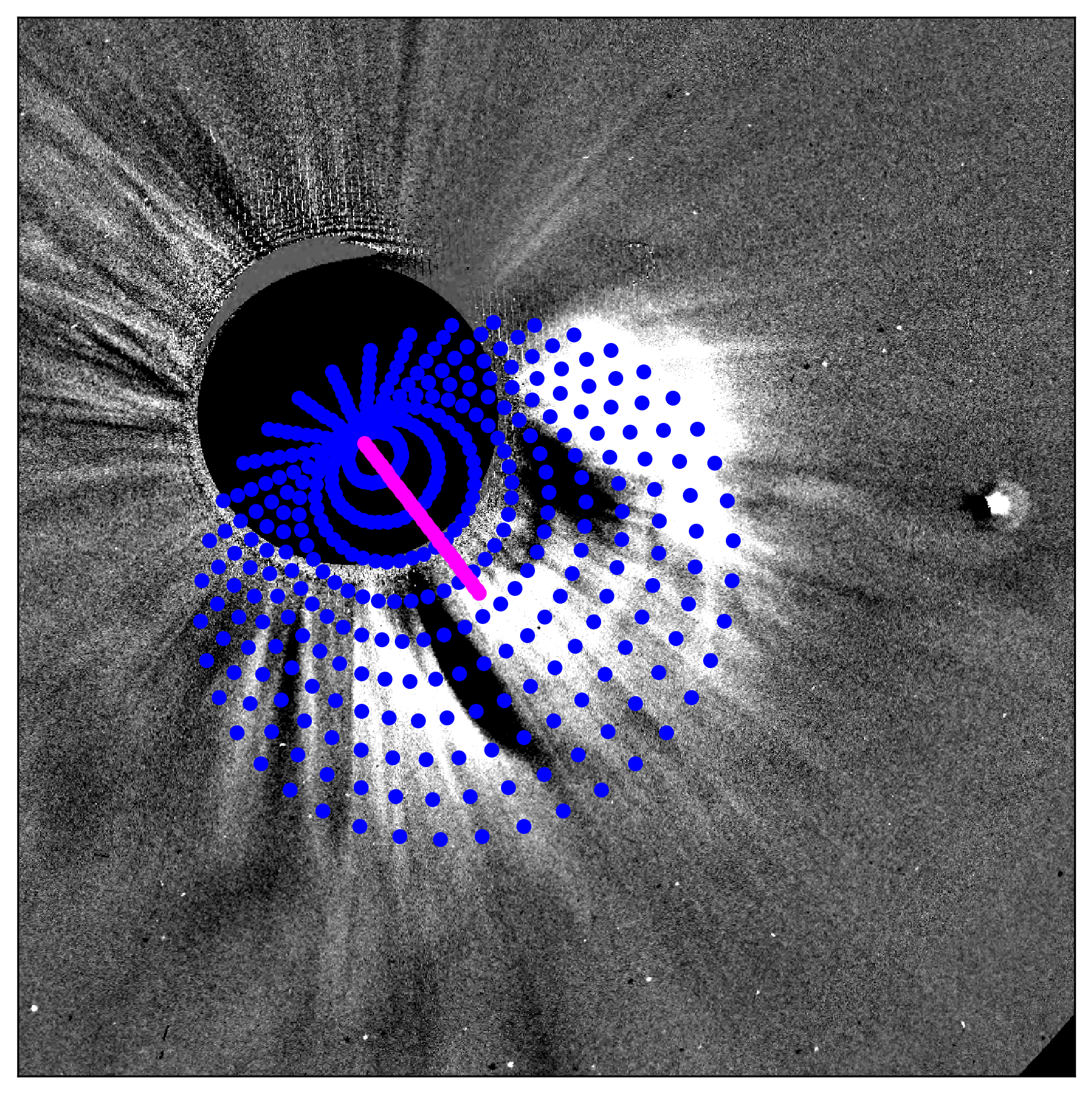}
		\caption{(a)}
		\label{}
	\end{subfigure}    
	\begin{subfigure}{0.49\columnwidth}   
		\centering
		\includegraphics[width=\linewidth]{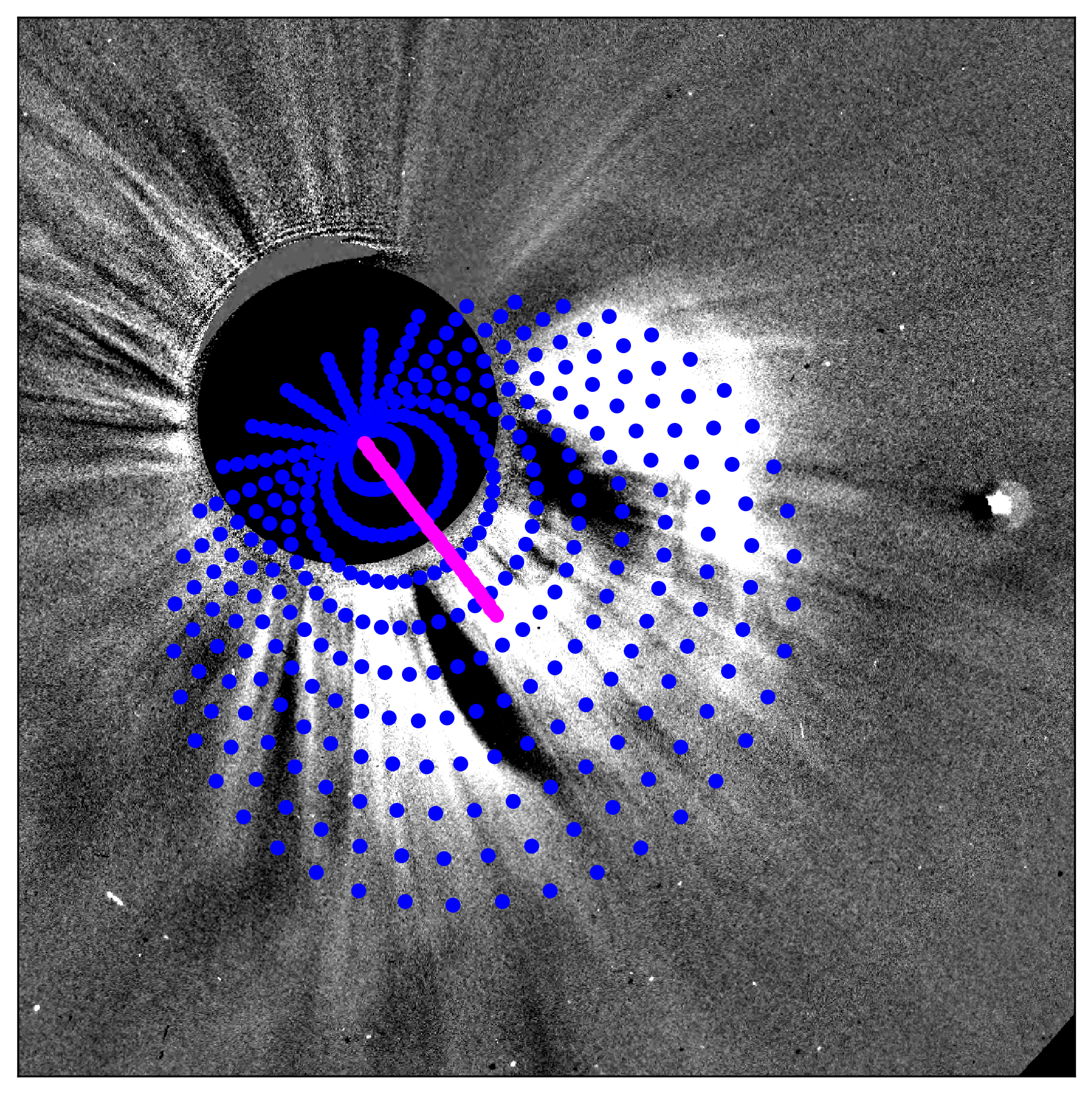}
		\caption{(b)}
		\label{}  
	\end{subfigure}
 \begin{subfigure}{0.49\columnwidth}
		\centering
		\includegraphics[width=\textwidth]{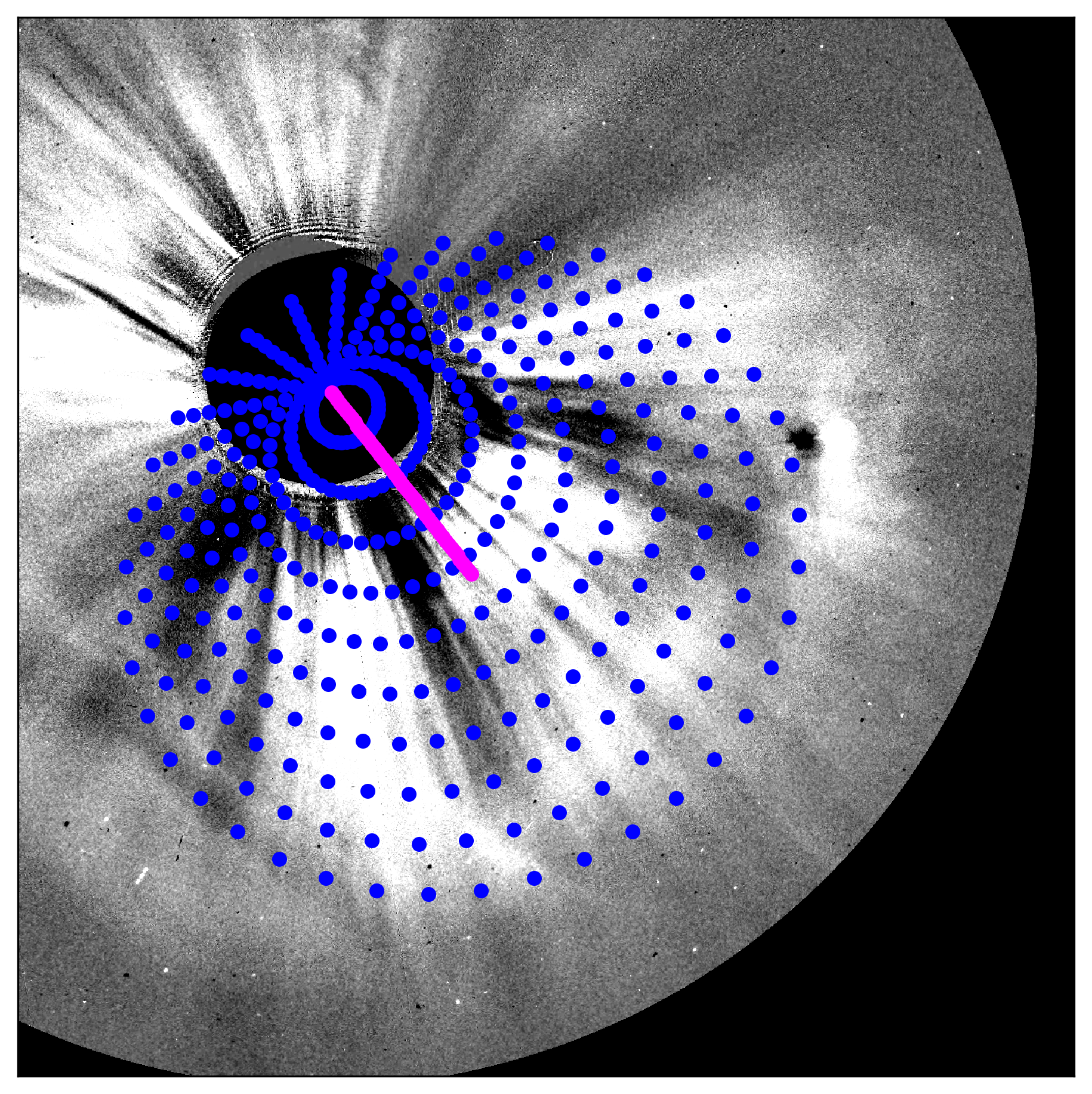}
		\caption{(c)}
		\label{}
	\end{subfigure}
    \begin{subfigure}{0.49\columnwidth}
		\centering
		\includegraphics[width=\textwidth]{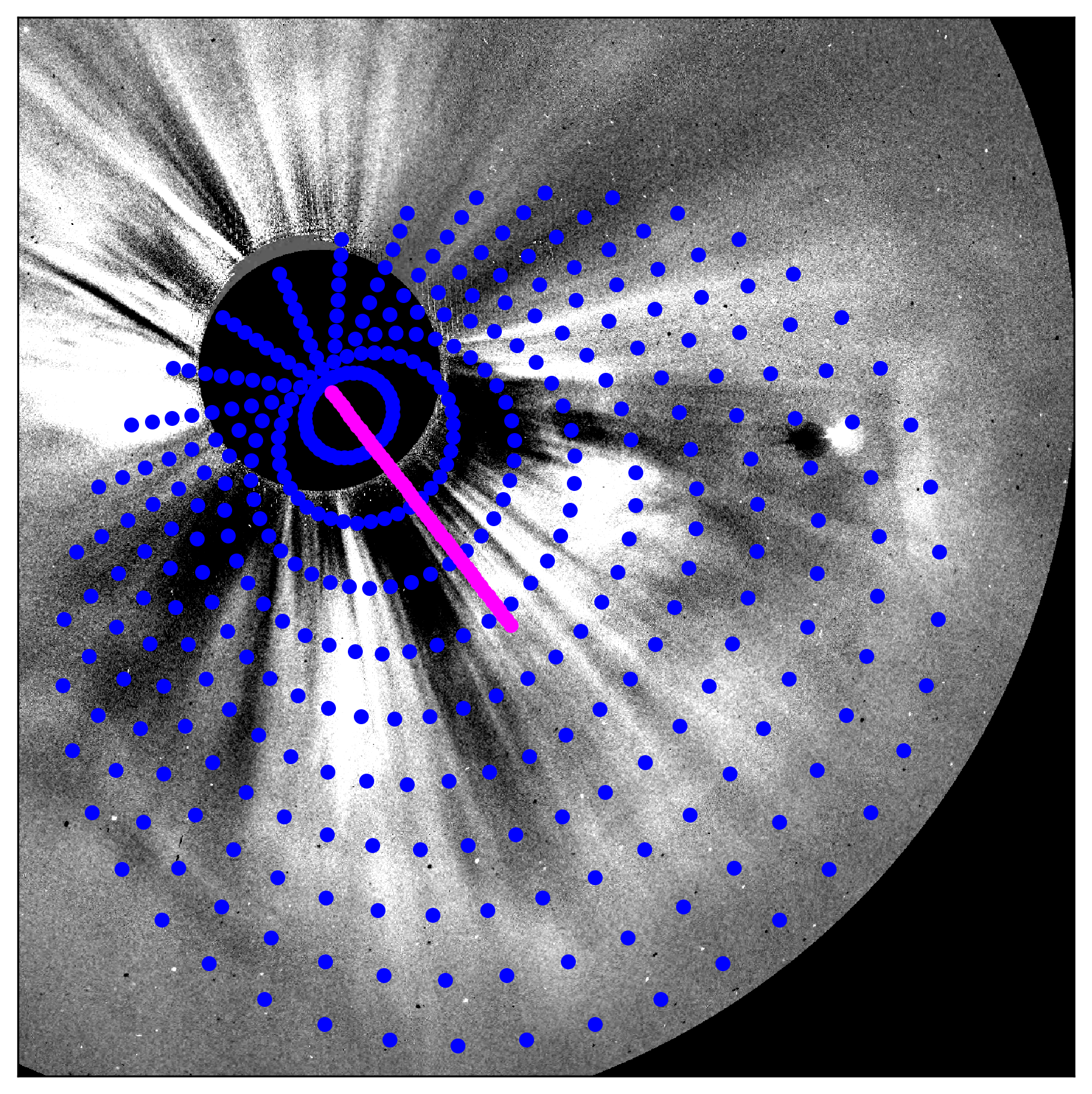}
		\caption{(d)}
		\label{}
	\end{subfigure}    
	\caption{LOS projections of the DIRECD cone (blue mesh) extended to (a) 6~$\rm{R_{sun}}$ at 05:38~UT , (b) 7~$\rm{R_{sun}}$ at 05:53~UT, (c) 10~$\rm{R_{sun}}$ at 08:53~UT and (d) 12~$\rm{R_{sun}}$ at 09:53~UT plotted on STEREO-A COR2 images. The pink line represents LOS projections of the cone central axes.}
	\label{COR_May}
\end{figure}

\begin{figure}
	\centering
	\includegraphics[width=0.5\textwidth]{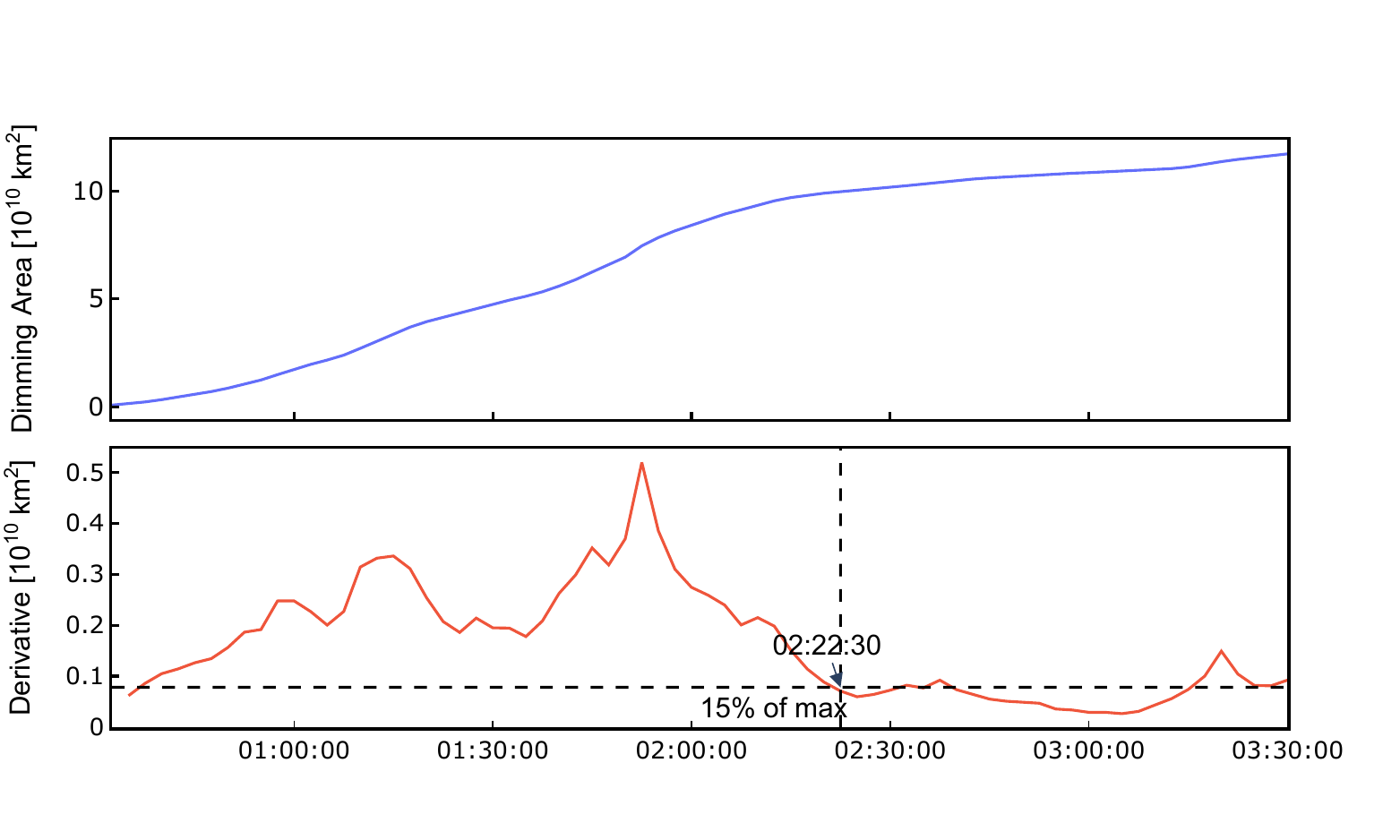}
	\caption{Expansion of dimming area $A_{t}$ (top panel) and its time derivative $dA_{t}/dt$ (bottom panel) over 3 hours for the 8 June, 2024 event. The end of the impulsive phase is defined as the time when the derivative of the dimming area curve, $dA_{t}/dt,$ has declined back to 15\% of its maximum value.}
	\label{area_der_june}
\end{figure}

\begin{figure}
	\centering
	\includegraphics[width=0.5\textwidth]{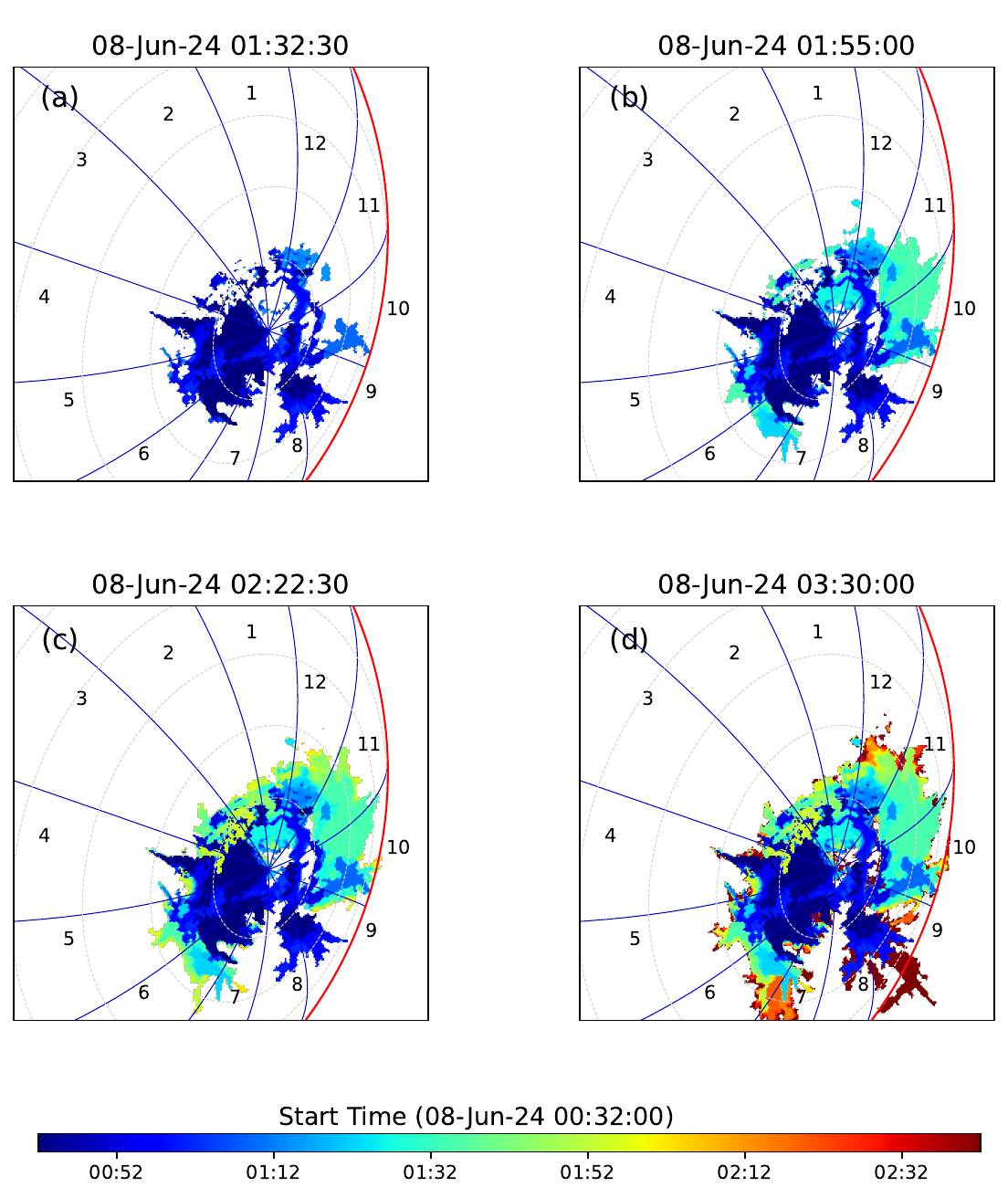}
	\caption{Dimming evolution for the 8 June 2024 CME at (a) 22 minutes before the maximum of the impulsive phase (01:32 UT), (b) at the maximum of impulsive phase (01:55 UT) , (c) at the end of the impulsive phase (02:22 UT) and (d) 3 hours after the start time of our analysis. The blue lines indicate 12 angular sectors and the colour bar shows when each dimming pixel was detected for the first time.}
	\label{dimming_expansion_june}
\end{figure}

To connect the structures of the eruption that we observe in the EUV with white-light coronagraph data, Figure~\ref{COR_May} shows line-of-sight (LOS) projections of the DIRECD cone extrapolated to 6~$R_{sun}$ at 05:38~UT (panel a), 05:53~UT (panel b), 08:53~UT (panel c) and 09:53~UT(panel d) capturing the moment when the CME is visible in STEREO-A COR2. 

The pink line shows the LOS projections of the central axis for the extended cones, pointing towards the centre of the CME bubble. As seen in Figure~\ref{COR_May}, the cone central line aligns closely with the inner parts of the CME bubble whereas the LOS projections of the extended 3D cones closely matches the edge of the CME structure observed at different time steps in COR2. This correspondence indicates that the connections between dimming and CME expansion are established by the end of the dimming's impulsive phase. Analysing the dimming evolution allowed us to extract key CME parameters early in its development, such as the 3D propagation direction and angular width, revealing a subsequent self-similar expansion. This finding is consistent with the studies by \citep{Chikunova2023} and \citep{jain2024coronal}, which demonstrated that dimming morphology closely reflects the inner part of 3D GCS CME reconstructions where both the DIRECD CME cone and the GCS croissant were reconstructed at the same time at the end of the dimming impulsive phase (the CME bubble was already seen in coronagraphs at this time). These results reinforce the idea that the extent of the dimming is related to CME propagation and leaves distinct footprints in the low corona, observable up to a certain height, as indicated by the parameters derived from DIRECD.

As both the DIRECD CME cone and the GCS croissant are reconstructed at the same time at the end of the dimming impulsive phase (the CME bubble is already seen in coronagraphs at this time), this supports the argument that the CME propagation is connected to the dimming and only leaves footprints in the low corona up to a limited height. These resemblances imply that the best-fit cone from the DIRECD method can be used not only to link the 2D dimming with the 3D CME bubble to provide an estimate of the early CME direction, but also to provide further information for improving estimations of GCS parameters using dimmings are signatures of CME propagation in the low corona. This can provide us with a comprehensive picture of CME evolution in the low corona (using DIRECD) to higher up in the heliosphere (using GCS or similar reconstructions).


\section{Event of June 8, 2024} 
On June 8, 2024 at 00:49 UT, active region AR3697 (previously AR3664) produced yet another intense solar flare (M9.7), ionizing the top of the Earth's atmosphere and causing shortwave radio blackouts across western Pacific Ocean. However, its impact on Earth was weaker and the associated CME did not cause a geomagnetic storm. The event was accompanied by a huge coronal dimming as observed from SDO and STEREO-A. For SDO, part of the dimming observed was very close to the limb and part of the eruption was off-limb. For STEREO-A, the dimming was observed on-disk, although also very close to the limb. Applying DIRECD, we will use STEREO-A on-disk dimming information.

We use a set of STEREO/EUVI (Extreme Ultraviolet Imager) images within the time range 00:30--03:30~UT, taking the first map as the base map (00:30~UT) and apply the same calibration steps and dimming detection as for May 2024 event, creating a series of base-difference images showing the absolute intensity-drop with respect to the pre-eruption state, with the threshold of $-40$ DN (chosen by analysing dimming development at different stages) and extracting the top 30\% of darkest pixels as seed pixels for the region growing algorithm.

\begin{figure}
	\begin{subfigure}{\columnwidth}
		\centering
	\includegraphics[width=0.8\textwidth]{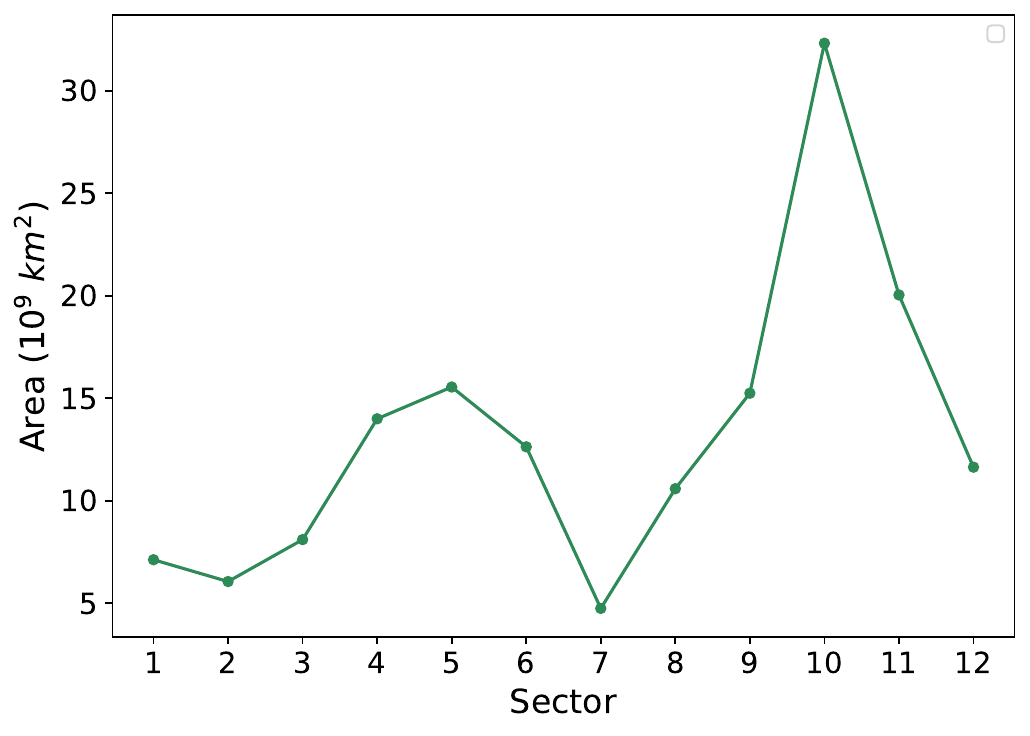}
		\caption{}
		\label{june_area_sector_end}
	\end{subfigure}    
	\begin{subfigure}{\columnwidth}    
		\includegraphics[width=0.9\textwidth]{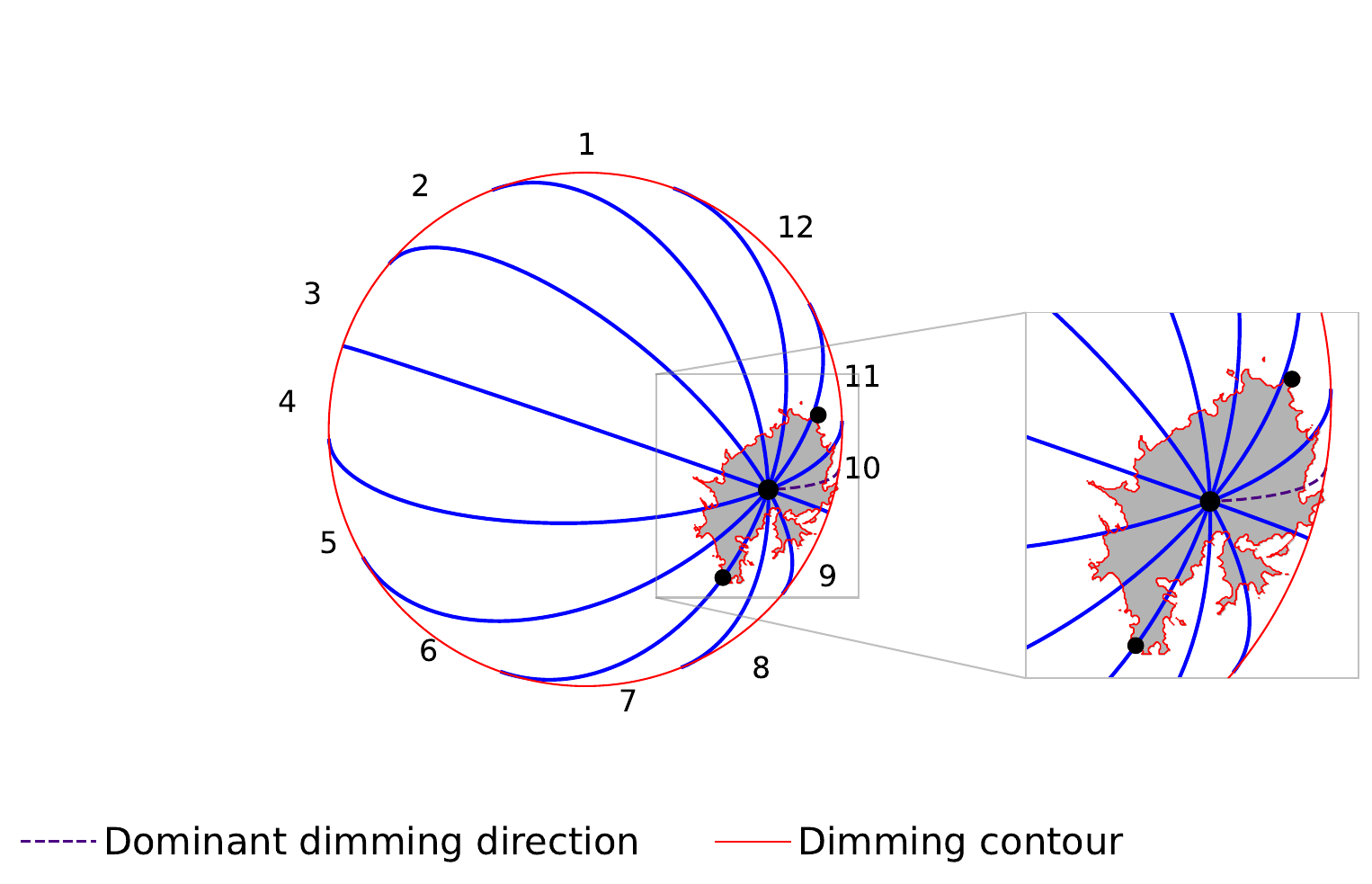}
		\caption{}
		\label{dominant_sector_june_end}  
	\end{subfigure}
	\caption{Dominant direction of dimming expansion at the end of the impulsive phase for 8 June 2024. Top: Dimming area in the different sectors (shown in the bottom panels), showing its maximum	at sector 10. Bottom: cumulative dimming pixel mask outlined by gray with red contours. The blue lines indicate the 12 sectors. The dashed purple line gives the sector of dominant dimming development. The small black dots show the dimming edges at the farthest edges from the source (one of them being in sector 11 and other being in sector 6) which are used to generate the 3D CME cones at different heights, associated widths and inclination angles.}
\end{figure}

Figure~\ref{area_der_june} shows the dimming area evolution over the 3 hours of interest. The end of the impulsive phase (15\% of maximum) is at 02:22:30~UT. The timing map shows that the dimming seems to be expanding radially from the source in all sectors (Figure~\ref{dimming_expansion_june}). We note that there is a partial off-limb dimming at the end of our three hour time period (panel d), however for DIRECD we use the dimming map at the end of impulsive phase (panel c), where the dimming expansion is on-disk.
As shown in Figure~\ref{june_area_sector_end}, the dominant dimming direction is in sector 10, illustrated by a dotted line in Figure~\ref{dominant_sector_june_end} along with the farthest dimming edges from the source (Edge 1 between sectors 6 and 7, and Edge 2 in sector 11), which we extend into 3D space to constrain the 3D-generated CME cones.

\begin{figure}
	\centering
	\includegraphics[width=0.5\textwidth]{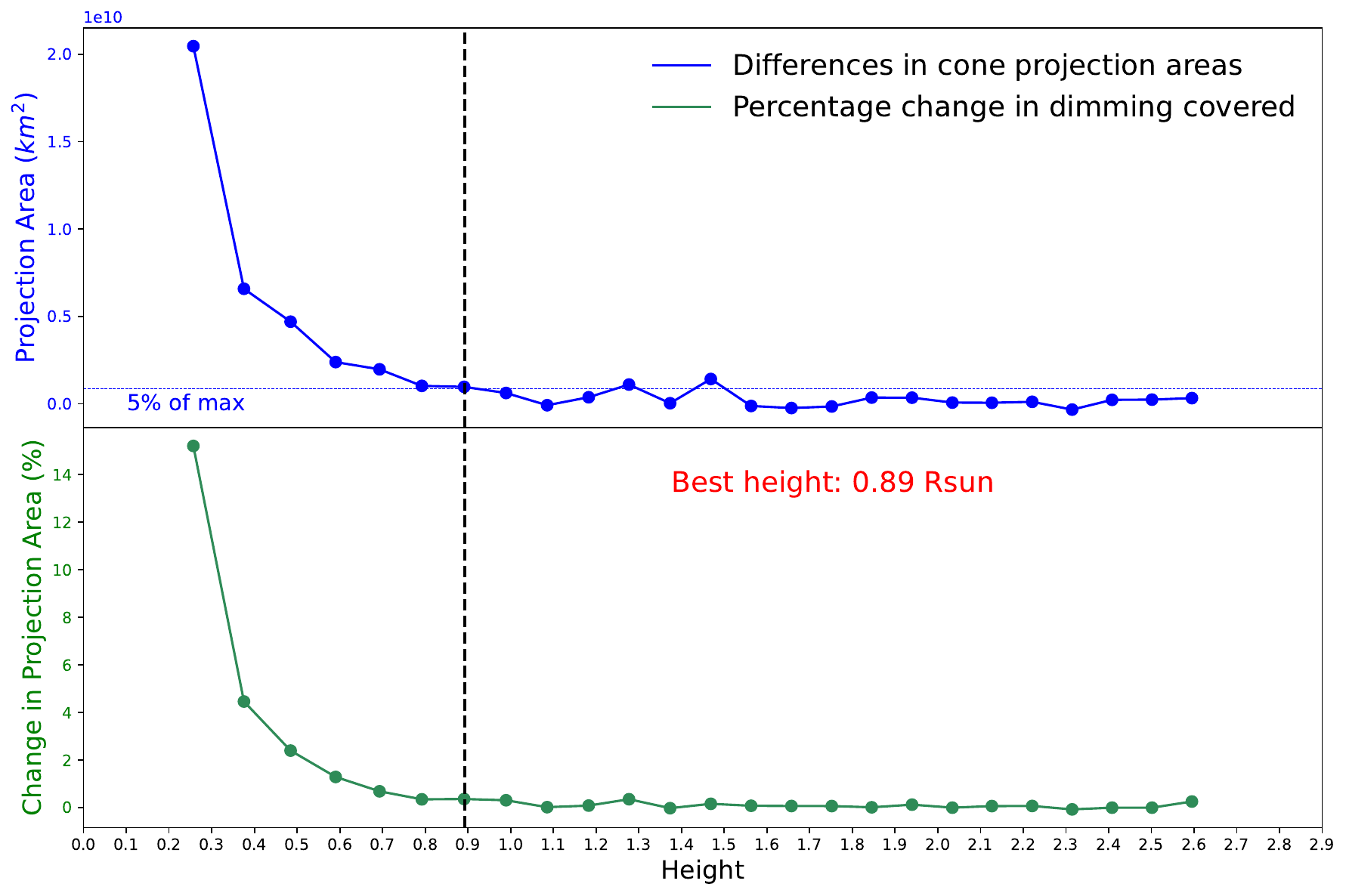}
	\caption{Same as Figure~\ref{best_height_may_1}, but for June 8, 2024.
		}
	\label{best_height_june_1}
\end{figure}

\begin{figure}
	\captionsetup[subfigure]{labelformat=empty}	
	\begin{subfigure}{0.5\columnwidth}
		\centering
    \includegraphics[width=\textwidth]{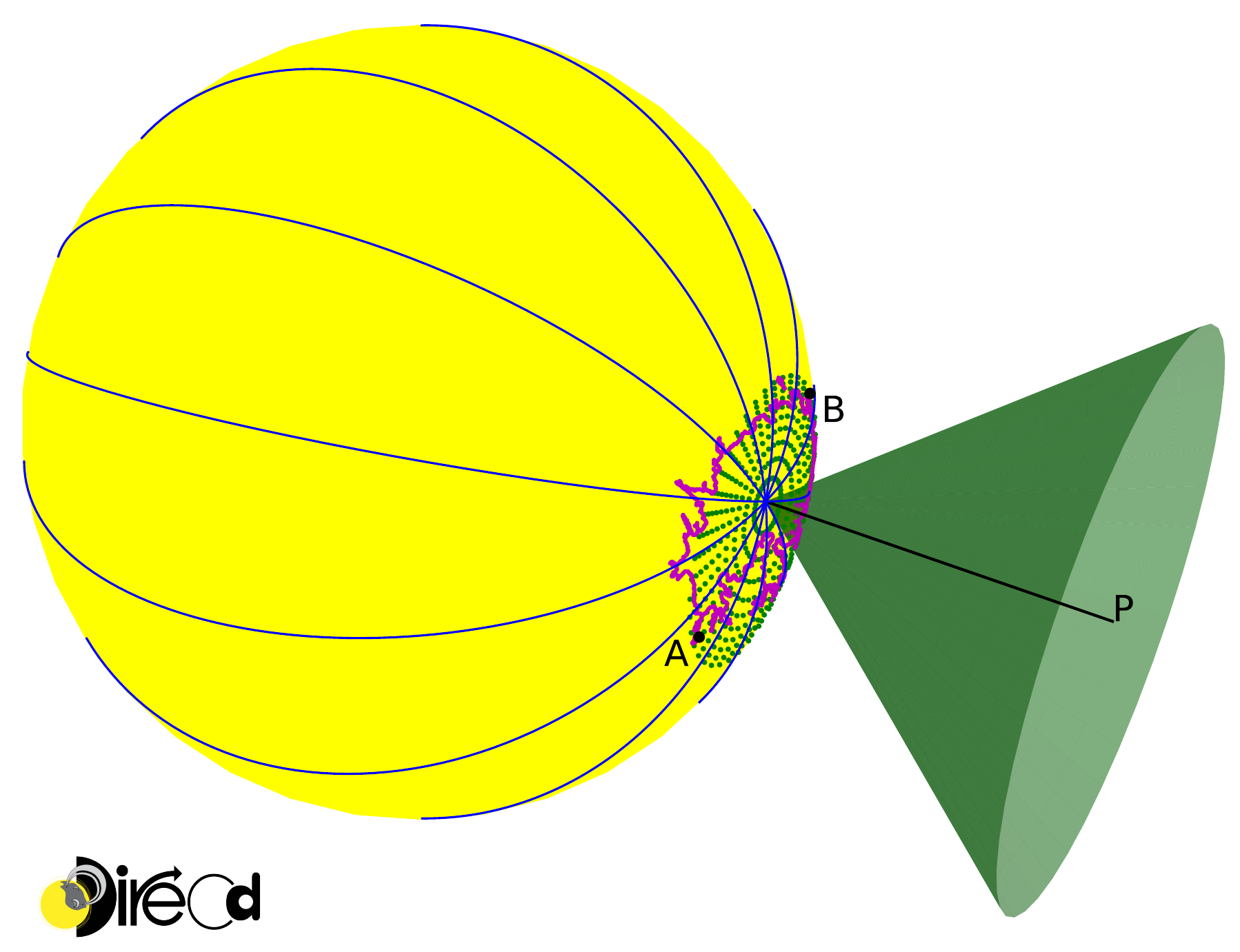}
		\caption{(a)}
		\label{best_fit_june}
	\end{subfigure}    
	\begin{subfigure}{0.45\columnwidth}    
		\centering
		\includegraphics[width=\textwidth]{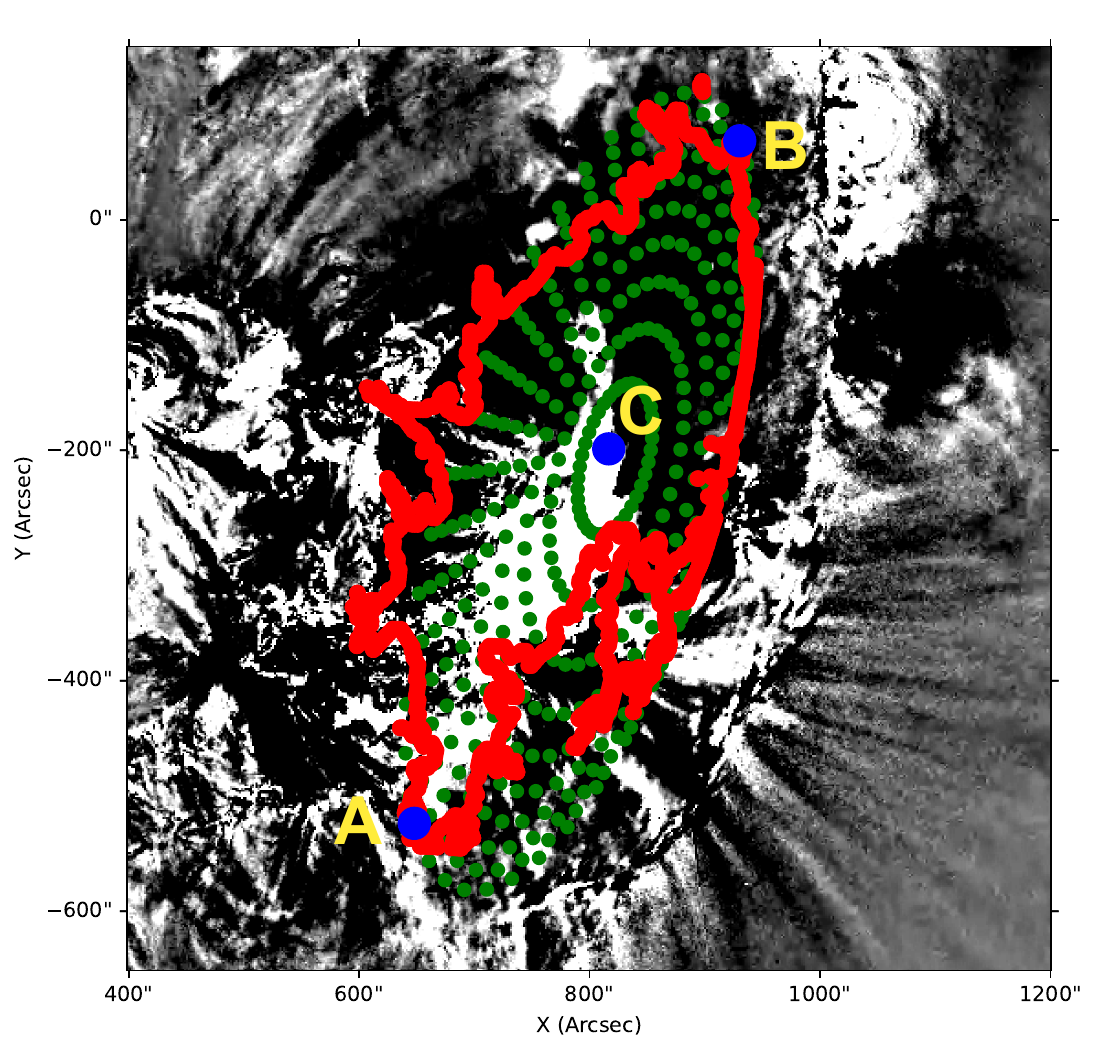}
		\caption{(b)}

		\label{STEREO_june}  
	\end{subfigure}
	\caption{(a) Application of the DIRECD method to relate the expansion of the coronal dimming to the early CME propagation. Magenta lines outline the dimming boundaries. Points A and B mark the largest North and South dimming extent.The best-fit	green cone
		has a height of 0.89 Rsun, width of 81$^\circ$ and inclination angle of 15.7$^\circ$ . Green dots indicate the orthogonal projections of the CME cone onto the solar surface. We require an edge of the cone base to be orthogonally projected to points A and B to match the dimming extent.(b) Dimming detection: 195Å EUVI/STEREO-A
		base-difference image together with the boundary of the identified dimming region (red) at the end of impulsive phase at 02:22 UT. Points A and B mark
		the largest North and South dimming extent.}
	\label{}
\end{figure}


Figure~\ref{june_projections} in Appendix~\ref{Appendix_A} shows an ensemble of twenty cones with heights ranging from $0.12-2.03$~$R_{sun}$, angular widths of $136.8-62.7^\circ$ and changing inclination angles from $52.5-11.1^\circ$ (Cols. 1 and 3) and and their orthogonal projections onto the solar sphere (Cols. 2 and 4). To determine the 3D CME cone that best matches the dimming geometry, we use the same criteria as for the May 8, 2024, event, as shown in Figure~\ref{best_height_june_1}.

We obtain that the best-fit 3D CME cone has a height of 0.89~$R_{sun}$ (where the CME still connects to the dimming), width of 81$^\circ$ and inclination angle of 15.7$^\circ$ (Figure \ref{best_fit_june}). Figure \ref{STEREO_june} shows a 195~\AA~STEREO-A EUVI base-difference image, along with the boundary of the identified dimming region (in red) at the end of the impulsive phase. The source is marked by point C. The largest extent of the dimming in the North and South direction is indicated by points A and B, respectively.

\begin{figure}  
	\centering      
	\includegraphics[width=0.35\textwidth] {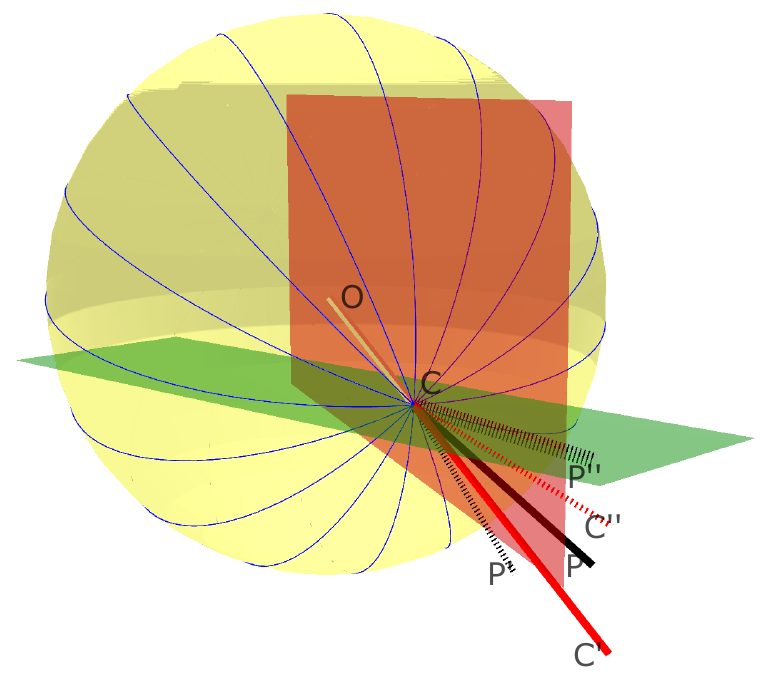}
	\caption{Same as Figure \ref{planes_may} but for June 8, 2024. The best fit cone is inclined 6.9$^\circ$ to the South (meridional plane) and 14.9$^\circ$ to the West(equatorial plane). }

	\label{planes_june}
\end{figure}

Figure \ref{planes_june} shows the best-fit cone inclination within the radial and meridional planes. We obtain that the 8 June 2024 is inclined from the radial direction by 6.9 $^\circ$  to the South (i.e, in the
meridional plane) and 14.9$^\circ$  to the West (in the equatorial plane).

\begin{figure}
	\captionsetup[subfigure]{labelformat=empty}	
	\begin{subfigure}{0.49\columnwidth}
		\centering
		\includegraphics[width=\textwidth]{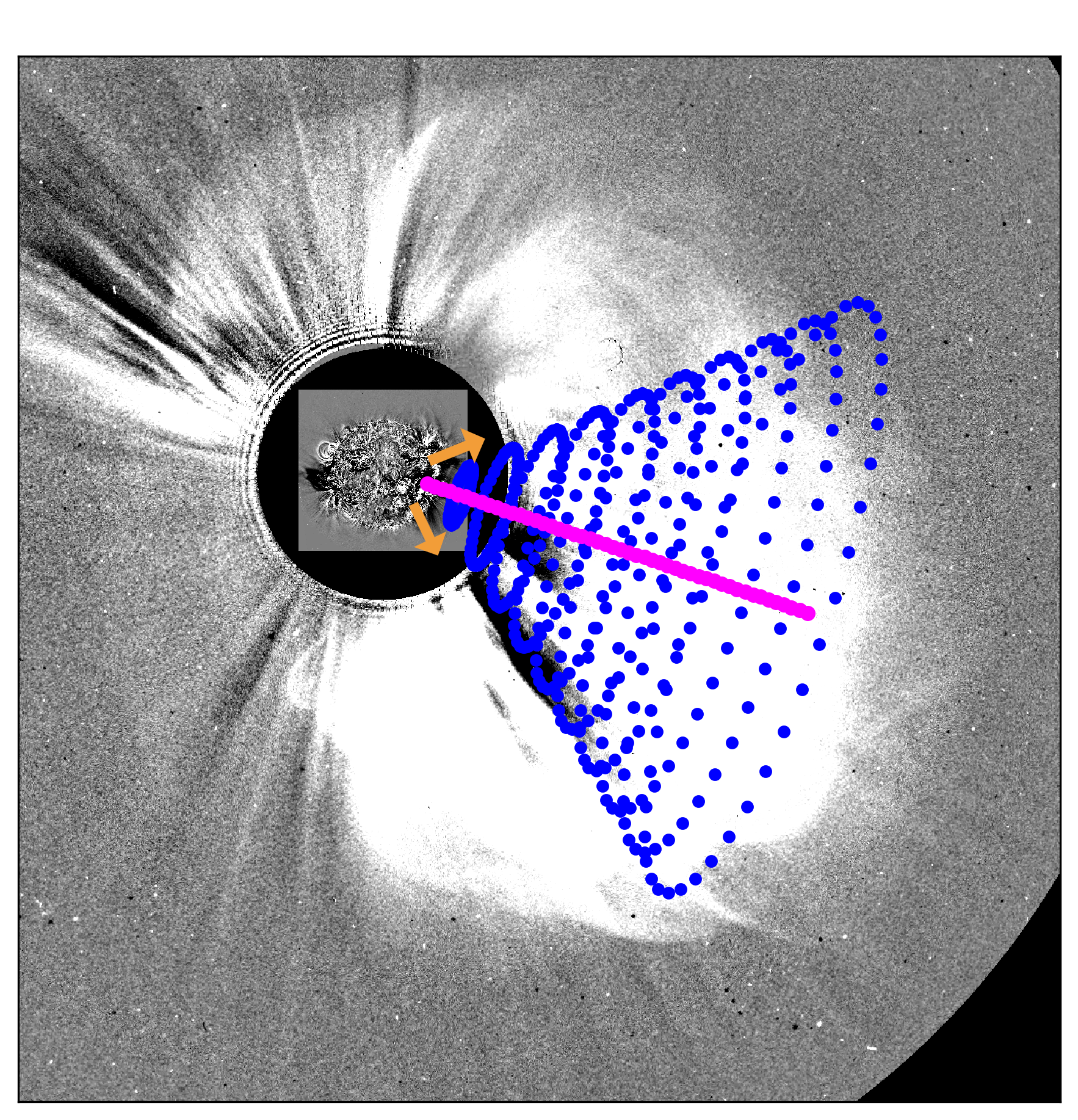}
		\caption{(a)}
		\label{}
	\end{subfigure}    
	\begin{subfigure}{0.49\columnwidth}   
		\centering
		\includegraphics[width=\linewidth]{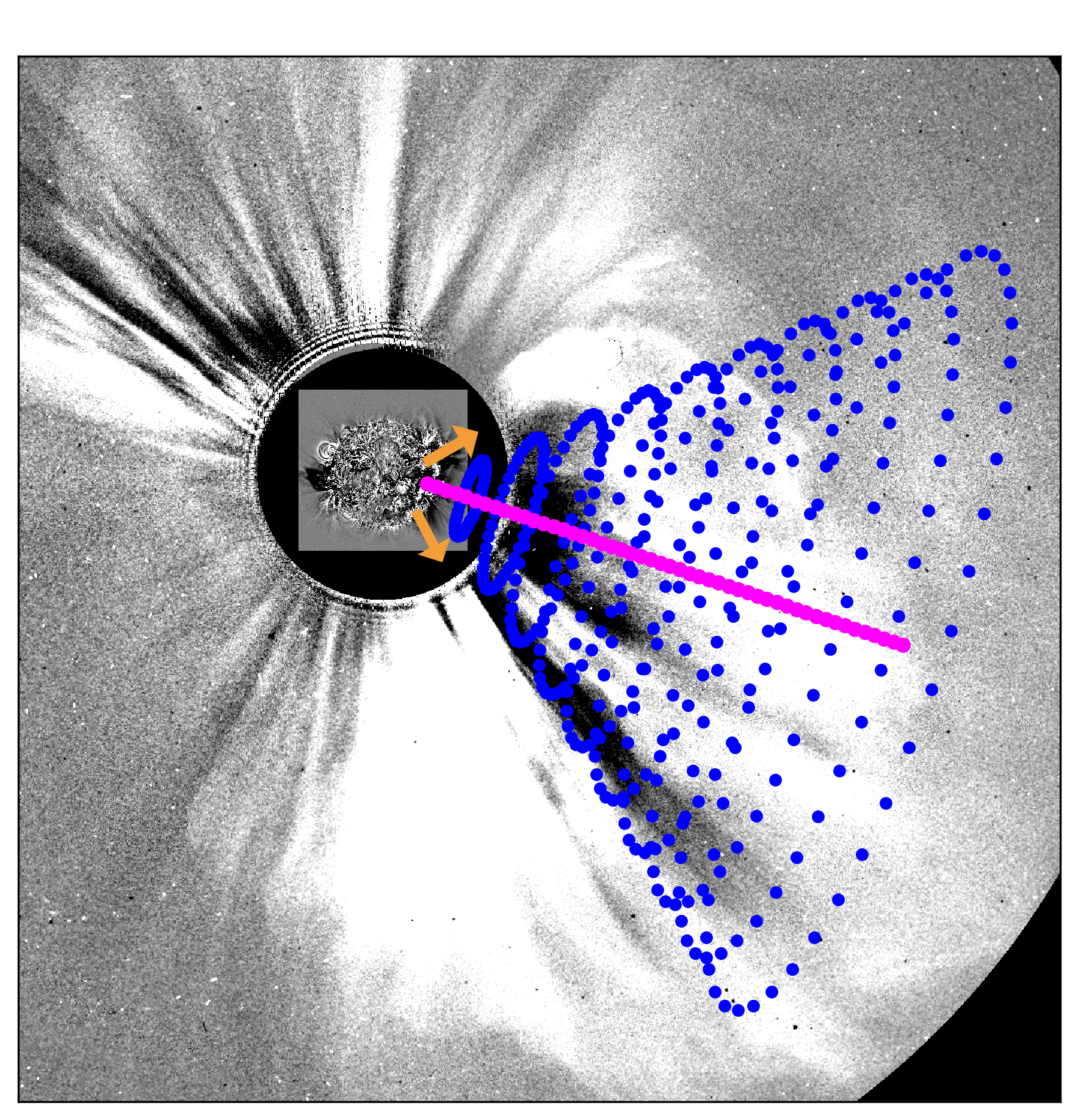}
		\caption{(b)}
		\label{}  
	\end{subfigure}
 
	\caption{LOS projections of the DIRECD cone extended to (a) 7~$\rm{R_{sun}}$ (blue mesh) on-top of a composite of 195~\AA~EUVI (02:22~UT) and COR2 (03:07~UT) and to (b) 9~$\rm{R_{sun}}$ (blue mesh) on-top of a composite of 195~\AA~EUVI (02:22~UT) and COR2 (04:07~UT). The pink line represents LOS projections of the cone central axes and the orange arrows show the direction of bubble from secondary dimming regions.}
	\label{COR_June_1}
\end{figure}

The best fit cone extrapolated to 7~$R_{sun}$ and 9~$R_{sun}$(keeping width and the inclination angle constant) and its LOS projections are plotted on both STEREO-A EUVI (02:22~UT) and COR2 (03:07~UT, 04:07~UT) images (Figure~\ref{COR_June_1}). 
As can be seen, the LOS projections of the cone's central axes are directed at the centre of the bubble which is captured by the core dimming. This confirms that dimmings serve as reliable indicators for deriving key CME parameters early in its development. It is to be noted that the secondary dimming regions further extend to the North and South of the initial core dimming. These correspond to the outer edges of CME bubble's white-light signature as observed in STEREO-A and are not fitted within DIRECD cone.

Remote dimming regions extend further to the North and South beyond the central dimming area. These regions, indicated by orange arrows in Figure \ref{COR_June_1}, are not accounted for in the DIRECD model but align with the outer boundaries of the CME bubble's white-light signature, as seen in the STEREO-A COR2 coronagraph image.

\section{Conclusions}

In this study, we estimated the early evolution and propagation of the CME direction with DIRECD for the severe May 8 and follow-up June 8, 2024 events from the expansion of the coronal dimming. It was shown that the link between dimming and CME expansion is established by the end of the dimming's impulsive phase, which allowed us to determine the key CME characteristics during the early CME evolution.

Using SDO/AIA 193~\AA~images for the May 8, 2024 event, we found that the CME expands close to radial, with an 
inclination angle of 7.7$^\circ$, angular width of 69.8$^\circ$ and a cone of height 0.81 $R_{sun}$, which was derived at the end of the dimming's impulsive phase, and for which the CME shows connections to the dimming and still leaves footprints in the low corona. The CME was inclined 7.6$^\circ$ to the North in the meridional plane and 1.1$^\circ$ to the East in the equatorial plane. For the follow-up June 8, 2024 event, we used STEREO/EUVI images for on-disk dimming and determined that the CME is inclined on 15.7$^\circ$, has an angular width of 81$^\circ$ and a height of 0.89~$R_{sun}$, for which the dimming is still connected to the CME. The CME was inclined 6.9$^\circ$ to the South in the meridional plane and 14.9$^\circ$ to the West in the equatorial plane.

Furthermore, we validated the 3D CME cone derived by DIRECD by correlating the CME properties derived in the low corona with white-light coronagraph data. We showed, that for both events the extended 3D CME cones, with the key parameters inferred from the EUV coronal dimmings, closely match the shape and edges of the CME structures simultaneously observed in the coronagraph and no CME deflection was observed upto the edge of STEREO-A field of view. By linking expanding dimming with the initial stages of CME development, this approach underscores coronal dimmings as valuable indicators of early CME evolution. It also suggests that the DIRECD method can be used not only to correlate 2D dimming with the 3D CME bubble but also to enhance GCS reconstructions with additional information about early CME propagation direction in the low corona.

The findings and outcomes of this study offer valuable insights into early CME evolution inferred from the coronal dimmings in low corona and DIRECD method proved to be efficient to provide the estimation of initial CME propagation direction and early CME parameters. Moreover, forecasting space weather involves using simulations like EUHFORIA \citep{pomoell2018euhforia} to model the behavior of CMEs, such as with the cone model, as they move through the heliosphere and approach Earth. For these forecasts to be reliable, the models must incorporate the most precise parameters available. A key parameter in this process is the CME’s propagation direction, defined by the insertion point's coordinates—longitude and latitude—at approximately 0.1 AU, as done in EUHFORIA. Using the DIRECD method could improve the accuracy of these key parameters at 0.1 AU, leading to more reliable predictions for space weather events. 


%
%



\begin{acknowledgements}
S.J., T.P. acknowledge support by the Russian Science Foundation under the project 23-22-00242, \url{https://rscf.ru/en/project/23-22-00242/}. G.C. and M.D. acknowledge the support by the Croatian Science Foundation under the project IP-2020-02-9893 (ICOHOSS). A.V. and M.D. acknowledge the support from the Austrian-Croatian Bilateral Scientific Project ”Analysis of solar eruptive phenomena from cradle to grave”. SDO data is courtesy of NASA/SDO and the AIA, and HMI science teams. The STEREO/SECCHI data are produced by an international consortium of the Naval Research Laboratory (USA), Lockheed Martin Solar and Astrophysics Lab (USA), NASA Goddard Space Flight centre (USA), Rutherford Appleton Laboratory (UK), University of Birmingham (UK), Max-Planck- Institut f\"ur Sonnenforschung (Germany), Centre Spatiale de Li\'ege (Belgium), Institut d'Optique Th\'eorique et Appliqu\'ee (France), and Institut d'Astrophysique Spatiale (France). We thank the referee for valuable comments on this study.
\end{acknowledgements}

\bibliographystyle{aa}
\bibliography{My_References}

\begin{thebibliography}{53}
\expandafter\ifx\csname natexlab\endcsname\relax\def\natexlab#1{#1}\fi

\bibitem[{{Aschwanden}(2017)}]{aschwanden2017global}
{Aschwanden}, M.~J. 2017, \apj, 847, 27

\bibitem[{{Attrill} {et~al.}(2006){Attrill}, {Nakwacki}, {Harra}, {Van Driel-Gesztelyi}, {Mandrini}, {Dasso}, \& {Wang}}]{attrill2006using}
{Attrill}, G., {Nakwacki}, M.~S., {Harra}, L.~K., {et~al.} 2006, \solphys, 238, 117

\bibitem[{{Baker} {et~al.}(2013){Baker}, {Li}, {Pulkkinen}, {Ngwira}, {Mays}, {Galvin}, \& {Simunac}}]{baker2013major}
{Baker}, D.~N., {Li}, X., {Pulkkinen}, A., {et~al.} 2013, Space Weather, 11, 585

\bibitem[{{Bewsher} {et~al.}(2008){Bewsher}, {Harrison}, \& {Brown}}]{bewsher2008relationship}
{Bewsher}, D., {Harrison}, R.~A., \& {Brown}, D.~S. 2008, \aap, 478, 897

\bibitem[{{Byrne} {et~al.}(2010){Byrne}, {Maloney}, {McAteer}, {Refojo}, \& {Gallagher}}]{byrne2010propagation}
{Byrne}, J.~P., {Maloney}, S.~A., {McAteer}, R.~T.~J., {Refojo}, J.~M., \& {Gallagher}, P.~T. 2010, Nature Communications, 1, 74

\bibitem[{{Cheng} {et~al.}(2017){Cheng}, {Guo}, \& {Ding}}]{Cheng2017}
{Cheng}, X., {Guo}, Y., \& {Ding}, M. 2017, Science China Earth Sciences, 60, 1383

\bibitem[{{Chikunova} {et~al.}(2020){Chikunova}, {Dissauer}, {Podladchikova}, \& {Veronig}}]{chikunova2020coronal}
{Chikunova}, G., {Dissauer}, K., {Podladchikova}, T., \& {Veronig}, A.~M. 2020, \apj, 896, 17

\bibitem[{{Chikunova} {et~al.}(2023){Chikunova}, {Podladchikova}, {Dissauer}, {Veronig}, {Dumbovi{\'c}}, {Temmer}, \& {Dickson}}]{Chikunova2023}
{Chikunova}, G., {Podladchikova}, T., {Dissauer}, K., {et~al.} 2023, \aap, 678, A166

\bibitem[{{Dissauer} {et~al.}(2019){Dissauer}, {Veronig}, {Temmer}, \& {Podladchikova}}]{Dissauer2019}
{Dissauer}, K., {Veronig}, A.~M., {Temmer}, M., \& {Podladchikova}, T. 2019, \apj, 874, 123

\bibitem[{{Dissauer} {et~al.}(2018{\natexlab{a}}){Dissauer}, {Veronig}, {Temmer}, {Podladchikova}, \& {Vanninathan}}]{Dissauer2018a}
{Dissauer}, K., {Veronig}, A.~M., {Temmer}, M., {Podladchikova}, T., \& {Vanninathan}, K. 2018{\natexlab{a}}, \apj, 855, 137

\bibitem[{{Dissauer} {et~al.}(2018{\natexlab{b}}){Dissauer}, {Veronig}, {Temmer}, {Podladchikova}, \& {Vanninathan}}]{Dissauer2018b}
{Dissauer}, K., {Veronig}, A.~M., {Temmer}, M., {Podladchikova}, T., \& {Vanninathan}, K. 2018{\natexlab{b}}, \apj, 863, 169

\bibitem[{Doherty {et~al.}(2004)Doherty, Coster, \& Murtagh}]{doherty2004space}
Doherty, P., Coster, A.~J., \& Murtagh, W. 2004, GPS Solutions, 8, 267 – 271, cited by: 78

\bibitem[{{Gonzalez} {et~al.}(1994){Gonzalez}, {Joselyn}, {Kamide}, {Kroehl}, {Rostoker}, {Tsurutani}, \& {Vasyliunas}}]{Gonzalez1994}
{Gonzalez}, W.~D., {Joselyn}, J.~A., {Kamide}, Y., {et~al.} 1994, \jgr, 99, 5771

\bibitem[{{Gopalswamy} {et~al.}(2009){Gopalswamy}, {Yashiro}, {Michalek}, {Stenborg}, {Vourlidas}, {Freeland}, \& {Howard}}]{gopalswamy2009soho}
{Gopalswamy}, N., {Yashiro}, S., {Michalek}, G., {et~al.} 2009, Earth Moon and Planets, 104, 295

\bibitem[{{Harrison} {et~al.}(2003){Harrison}, {Bryans}, {Simnett}, \& {Lyons}}]{harrison2003coronal}
{Harrison}, R.~A., {Bryans}, P., {Simnett}, G.~M., \& {Lyons}, M. 2003, \aap, 400, 1071

\bibitem[{{Harrison} \& {Lyons}(2000)}]{harrison2000spectroscopic}
{Harrison}, R.~A. \& {Lyons}, M. 2000, \aap, 358, 1097

\bibitem[{Hayakawa {et~al.}(2024)Hayakawa, Ebihara, Mishev, Koldobskiy, Kusano, Bechet, Yashiro, Iwai, Shinbori, Mursula, {et~al.}}]{hayakawa2024solar}
Hayakawa, H., Ebihara, Y., Mishev, A., {et~al.} 2024, arXiv preprint arXiv:2407.07665

\bibitem[{{Hudson} {et~al.}(1996){Hudson}, {Acton}, \& {Freeland}}]{hudson1996long}
{Hudson}, H.~S., {Acton}, L.~W., \& {Freeland}, S.~L. 1996, \apj, 470, 629

\bibitem[{Jain {et~al.}(2024)Jain, Podladchikova, Chikunova, Dissauer, \& Veronig}]{jain2024coronal}
Jain, S., Podladchikova, T., Chikunova, G., Dissauer, K., \& Veronig, A.~M. 2024, Astronomy \& Astrophysics, 683, A15

\bibitem[{Kaiser {et~al.}(2008)Kaiser, Kucera, Davila, St~Cyr, Guhathakurta, \& Christian}]{kaiser2008stereo}
Kaiser, M.~L., Kucera, T., Davila, J., {et~al.} 2008, Space Science Reviews, 136, 5

\bibitem[{{Krista} \& {Reinard}(2017)}]{krista2017statistical}
{Krista}, L.~D. \& {Reinard}, A.~A. 2017, \apj, 839, 50

\bibitem[{Lemen {et~al.}(2012)Lemen, Title, Akin, Boerner, Chou, Drake, Duncan, Edwards, Friedlaender, Heyman, {et~al.}}]{lemen2012atmospheric}
Lemen, J.~R., Title, A.~M., Akin, D.~J., {et~al.} 2012, Solar Physics, 275, 17

\bibitem[{{Liu} {et~al.}(2010){Liu}, {Thernisien}, {Luhmann}, {Vourlidas}, {Davies}, {Lin}, \& {Bale}}]{liu2010reconstructing}
{Liu}, Y., {Thernisien}, A., {Luhmann}, J.~G., {et~al.} 2010, \apj, 722, 1762

\bibitem[{{L{\'o}pez} {et~al.}(2017){L{\'o}pez}, {Hebe Cremades}, {Nuevo}, {Balmaceda}, \& {V{\'a}squez}}]{lopez2017mass}
{L{\'o}pez}, F.~M., {Hebe Cremades}, M., {Nuevo}, F.~A., {Balmaceda}, L.~A., \& {V{\'a}squez}, A.~M. 2017, \solphys, 292, 6

\bibitem[{Lugaz {et~al.}(2017)Lugaz, Temmer, Wang, \& Farrugia}]{lugaz2017interaction}
Lugaz, N., Temmer, M., Wang, Y., \& Farrugia, C.~J. 2017, Solar Physics, 292, 1

\bibitem[{{Mason} {et~al.}(2016){Mason}, {Woods}, {Webb}, {Thompson}, {Colaninno}, \& {Vourlidas}}]{mason2016relationship}
{Mason}, J.~P., {Woods}, T.~N., {Webb}, D.~F., {et~al.} 2016, \apj, 830, 20

\bibitem[{{Michalek} {et~al.}(2009){Michalek}, {Gopalswamy}, \& {Yashiro}}]{michalek2009expansion}
{Michalek}, G., {Gopalswamy}, N., \& {Yashiro}, S. 2009, \solphys, 260, 401

\bibitem[{{Miklenic} {et~al.}(2011){Miklenic}, {Veronig}, {Temmer}, {M{\"o}stl}, \& {Biernat}}]{miklenic2011coronal}
{Miklenic}, C., {Veronig}, A.~M., {Temmer}, M., {M{\"o}stl}, C., \& {Biernat}, H.~K. 2011, \solphys, 273, 125

\bibitem[{M{\"u}ller {et~al.}(2017)M{\"u}ller, Nicula, Felix, Verstringe, Bourgoignie, Csillaghy, Berghmans, Jiggens, Garc{\'\i}a-Ortiz, Ireland, {et~al.}}]{muller2017jhelioviewer}
M{\"u}ller, D., Nicula, B., Felix, S., {et~al.} 2017, Astronomy \& Astrophysics, 606, A10

\bibitem[{Parker \& Linares(2024)}]{parker2024satellite}
Parker, W.~E. \& Linares, R. 2024, Journal of Spacecraft and Rockets, 1

\bibitem[{Pesnell {et~al.}(2012)Pesnell, Thompson, \& Chamberlin}]{pesnell2012solar}
Pesnell, W.~D., Thompson, B.~J., \& Chamberlin, P. 2012, The solar dynamics observatory (SDO) (Springer)

\bibitem[{{Podladchikova} {et~al.}(2018){Podladchikova}, {Petrukovich}, \& {Yermolaev}}]{Podladchikova2018}
{Podladchikova}, T., {Petrukovich}, A., \& {Yermolaev}, Y. 2018, Journal of Space Weather and Space Climate, 8, A22

\bibitem[{{Podladchikova} {et~al.}(2019){Podladchikova}, {Veronig}, {Dissauer}, {Temmer}, \& {Podladchikova}}]{Podladchikova2019}
{Podladchikova}, T., {Veronig}, A.~M., {Dissauer}, K., {Temmer}, M., \& {Podladchikova}, O. 2019, \apj, 877, 68

\bibitem[{{Podladchikova} \& {Petrukovich}(2012)}]{Podladchikova2012}
{Podladchikova}, T.~V. \& {Petrukovich}, A.~A. 2012, Space Weather, 10, S07001

\bibitem[{Pomoell \& Poedts(2018)}]{pomoell2018euhforia}
Pomoell, J. \& Poedts, S. 2018, Journal of Space Weather and Space Climate, 8, A35

\bibitem[{{Qiu} \& {Cheng}(2017)}]{qiu2017gradual}
{Qiu}, J. \& {Cheng}, J. 2017, \apjl, 838, L6

\bibitem[{{Reinard} \& {Biesecker}(2009)}]{reinard2009relationship}
{Reinard}, A.~A. \& {Biesecker}, D.~A. 2009, \apj, 705, 914

\bibitem[{{Rodr{\'\i}guez G{\'o}mez} {et~al.}(2020){Rodr{\'\i}guez G{\'o}mez}, {Podladchikova}, {Veronig}, {Ruzmaikin}, {Feynman}, \& {Petrukovich}}]{gomez2020clustering}
{Rodr{\'\i}guez G{\'o}mez}, J.~M., {Podladchikova}, T., {Veronig}, A., {et~al.} 2020, \apj, 899, 47

\bibitem[{Ronca {et~al.}(2024)Ronca, Chikunova, Dissauer, Podladchikova, \& Veronig}]{ronca2024recoverycoronaldimmings}
Ronca, G.~M., Chikunova, G., Dissauer, K., Podladchikova, T., \& Veronig, A.~M. 2024, Recovery of coronal dimmings

\bibitem[{{Sandford}(1999)}]{sandford1999impact}
{Sandford}, W.~H. 1999, Journal of Navigation, 52, 42

\bibitem[{{Schwenn} {et~al.}(2005){Schwenn}, {dal Lago}, {Huttunen}, \& {Gonzalez}}]{schwenn2005association}
{Schwenn}, R., {dal Lago}, A., {Huttunen}, E., \& {Gonzalez}, W.~D. 2005, Annales Geophysicae, 23, 1033

\bibitem[{Scolini {et~al.}(2020)Scolini, Chan{\'e}, Temmer, Kilpua, Dissauer, Veronig, Palmerio, Pomoell, Dumbovi{\'c}, Guo, {et~al.}}]{scolini2020cme}
Scolini, C., Chan{\'e}, E., Temmer, M., {et~al.} 2020, The Astrophysical Journal Supplement Series, 247, 21

\bibitem[{Spogli {et~al.}(2024)Spogli, Alberti, Bagiacchi, Cafarella, Cesaroni, Cianchini, Coco, Di~Mauro, Ghidoni, Giannattasio, {et~al.}}]{spogli2024effects}
Spogli, L., Alberti, T., Bagiacchi, P., {et~al.} 2024, Annals of Geophysics, 67, PA218

\bibitem[{{Sterling} \& {Hudson}(1997)}]{sterling1997yohkoh}
{Sterling}, A.~C. \& {Hudson}, H.~S. 1997, \apjl, 491, L55

\bibitem[{{The SunPy Community} {et~al.}(2020){The SunPy Community}, Barnes, Bobra, Christe, Freij, Hayes, Ireland, Mumford, Perez-Suarez, Ryan, Shih, Chanda, Glogowski, Hewett, Hughitt, Hill, Hiware, Inglis, Kirk, Konge, Mason, Maloney, Murray, Panda, Park, Pereira, Reardon, Savage, Sipőcz, Stansby, Jain, Taylor, Yadav, Rajul, \& Dang}]{sunpy_community2020}
{The SunPy Community}, Barnes, W.~T., Bobra, M.~G., {et~al.} 2020, The Astrophysical Journal, 890, 68

\bibitem[{{Thernisien}(2011)}]{thernisien2011implementation}
{Thernisien}, A. 2011, \apjs, 194, 33

\bibitem[{{Thernisien} {et~al.}(2006){Thernisien}, {Howard}, \& {Vourlidas}}]{thernisien2006modeling}
{Thernisien}, A.~F.~R., {Howard}, R.~A., \& {Vourlidas}, A. 2006, \apj, 652, 763

\bibitem[{{Thompson} {et~al.}(1998){Thompson}, {Plunkett}, {Gurman}, {Newmark}, {St. Cyr}, \& {Michels}}]{thompson1998soho}
{Thompson}, B.~J., {Plunkett}, S.~P., {Gurman}, J.~B., {et~al.} 1998, \grl, 25, 2465

\bibitem[{{Tsurutani} \& {Lakhina}(2014)}]{tsurutani2014extreme}
{Tsurutani}, B.~T. \& {Lakhina}, G.~S. 2014, \grl, 41, 287

\bibitem[{Vennerstrom {et~al.}(2016)Vennerstrom, Lefevre, Dumbovi{\'c}, Crosby, Malandraki, Patsou, Clette, Veronig, Vr{\v{s}}nak, Leer, {et~al.}}]{vennerstrom2016extreme}
Vennerstrom, S., Lefevre, L., Dumbovi{\'c}, M., {et~al.} 2016, Solar physics, 291, 1447

\bibitem[{{Veronig} {et~al.}(2018){Veronig}, {Podladchikova}, {Dissauer}, {Temmer}, {Seaton}, {Long}, {Guo}, {Vr{\v{s}}nak}, {Harra}, \& {Kliem}}]{Veronig2018_Genesis}
{Veronig}, A.~M., {Podladchikova}, T., {Dissauer}, K., {et~al.} 2018, \apj, 868, 107

\bibitem[{Watanabe {et~al.}(2012)Watanabe, Masuda, \& Segawa}]{watanabe2012hinode}
Watanabe, K., Masuda, S., \& Segawa, T. 2012, Solar Physics, 279, 317

\bibitem[{{Zhukov} \& {Auch{\`e}re}(2004)}]{zhukov2004nature}
{Zhukov}, A.~N. \& {Auch{\`e}re}, F. 2004, \aap, 427, 705

\end{thebibliography}

\begin{appendix}

\section{Ensemble of cones}\label{Appendix_A}
\begin{figure}[h]  
	\centering      
	\includegraphics[width=0.4\textwidth] {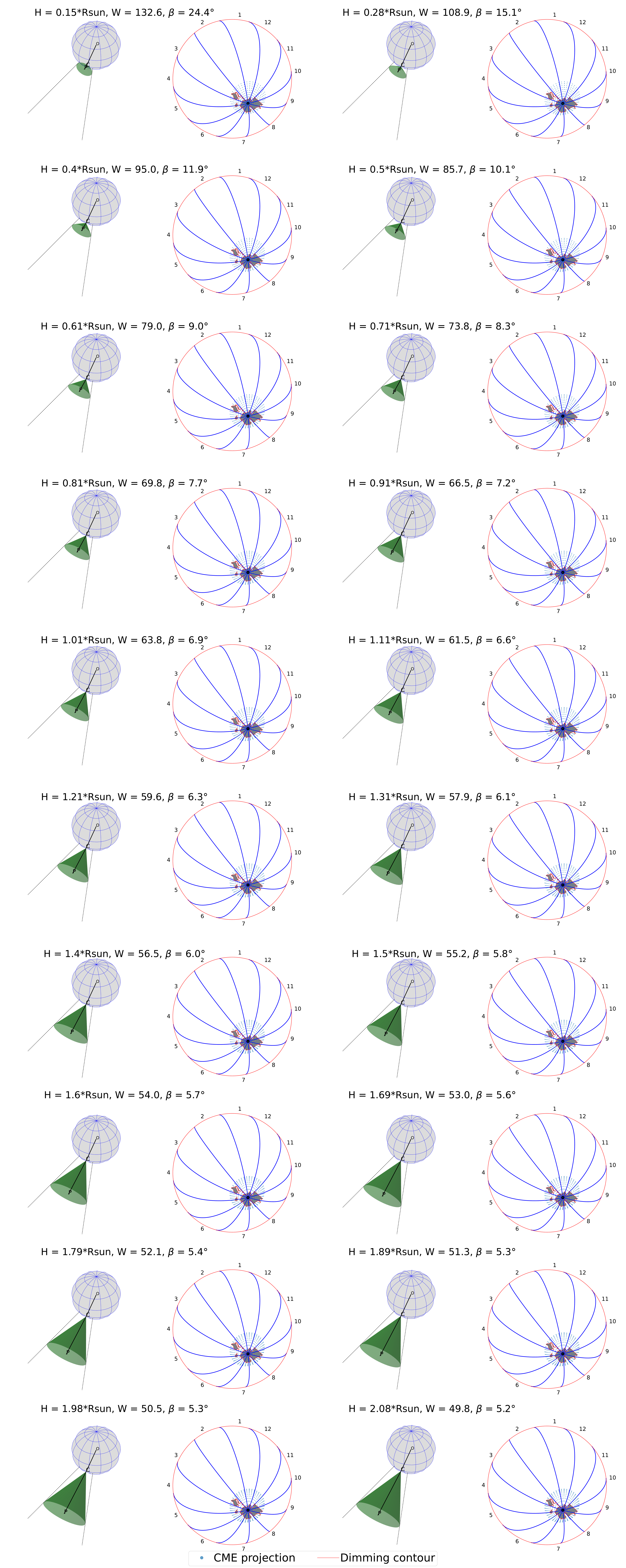}
	\caption{3D CME cones at heights of 0.15--2.08~$R_{sun}$ with widths of 132.6--49.8$^\circ$ and inclination angles of 24.4--5.2$^\circ$ for 8 May 2024 (columns 1 and 3), and their orthogonal projections onto the solar sphere (columns 2 and 4) bounded by the dimming edges.
	} 
	\label{may_projections}
\end{figure}

\begin{figure}[h]  
	\centering   
    \vspace{1.2cm}
	\includegraphics[width=0.4\textwidth] {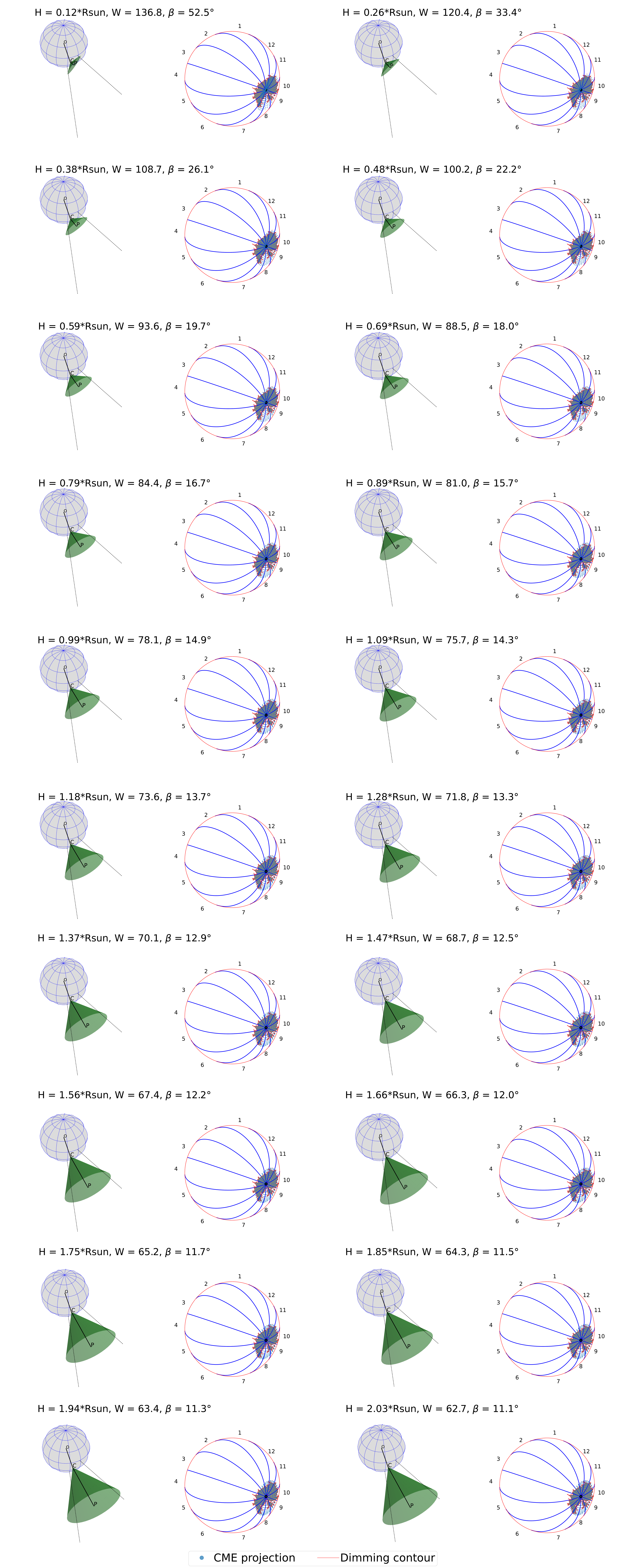}
	\caption{Same as Figure~\ref{may_projections}, but for June 8, 2024.
	} 
	\label{june_projections}
\end{figure}
\clearpage
	\section{Percentage of relative change of the dimming area inside the projection for the consecutive cones}\label{Appendix_B}
	
In this section, we present a detailed explanation of the criterion used to derive the 3D CME cone, which matches best to the dimming geometry, presented by Equation~\ref{criteria_2}.

With increase in the cone height, Equation~\ref{criteria_2} is monotonically decreasing, which means that denominator $S_{k}$ decreases for each next step, and $S_{k}~<~S_{k+1}$.
	
Let us consider the introduced criterion at two consecutive steps
	\begin{equation}\label{criteria_2_2}
		\frac{S_k - S_{k+1}}{S_k} - \frac{S_{k-1} - S_k}{S_{k-1}} < 0
\end{equation}
As follows from Equation~\ref{criteria_2_2}, a value of each further shrinking of dimming area inside the projection also decreases, so $S_{k}-S_{k+1} < S_{k-1} - S_{k}$.  This indicates the removal of insignificant or possibly random outer fragments of the dimming, which may not be in direct contact with the main dimming area (such as isolated islands).
As also follows from Equation~\ref{criteria_2_2}, the ratio of the further shrinking of the dimming area inside the projection to the previous one is smaller than the ratio of the areas inside the projections (Equation~\ref{criteria_2_3})
\begin{equation}\label{criteria_2_3}
		\frac{S_k - S_{k+1}}{S_{k-1} - S_k} - \frac{S_k}{S_{k-1}} < 0
\end{equation}
However, with a further increase of the cone height, a situation may change sharply, and the sign at a step $k_0+1$ is changed to the opposite one
\begin{equation*}
		\frac{S_{k_0} - S_{k_0 + 1}}{S_{k_0}} - \frac{S_{k_0 - 1} - S_{k_0}}{S_{k_0 - 1}} > 0 
\end{equation*}
Or
\begin{equation}\label{criteria_2_4}
		\frac{S_{k_0} - S_{k_0 + 1}}{S_{k_0 - 1} - S_{k_0}} - \frac{S_{k_0}}{S_{k_0 - 1}} > 0
\end{equation}
	
The non-equality in Equation~\ref{criteria_2_4} changes the sign, as on the step $k_0+1$ the area of dimming inside the projection $S_{k_0+1}$ decreased more than on a previous steps. This leads to a strong increase of the numerator of ratio $\frac{S_{k_0} - S_{k_0 + 1}}{S_{k_0 - 1} - S_{k_0}}$ which is even more, than the ratio of just areas.
	
The sharp decrease in the dimming area inside the projection is due to a significant increase in the area of contact between the projection perimeter and the selected dimming at step $k_0+1$ compared to the previous steps. A further increase in height and a corresponding reduction in projection obviously leads to an undue reduction in the dimming area and distortion of the results. Therefore, we use condition~\ref{criteria_2_4} as a criterion determining the choice of the best-fit 3D cone and associated CME parameters.
	
\end{appendix}

\end{document}